\documentclass[11pt, oneside]{article}   	
\usepackage[margin=1in]{geometry}               
\usepackage{graphicx}				
										
\usepackage{amssymb}
\usepackage{physics}
\usepackage{amsmath}
\usepackage[dvips]{epsfig}

\usepackage{amsfonts}
\usepackage{subfig}

\usepackage{mathrsfs}

\usepackage{xcolor}

\usepackage{bm}

\usepackage[title]{appendix}

\usepackage{cite}

\usepackage{authblk}

\def\cchi{\raise2pt\hbox{$\chi$}} 

\title{\bf{Qubit Lattice Algorithms based on the Schrodinger-Dirac representation of Maxwell Equations and their Extensions}}
\author {George Vahala ${ }^{1}$, Min Soe ${ }^{2}$, Efstratios Koukoutsis ${ }^{3}$, Kyriakos Hizanidis ${ }^{3}$, Linda Vahala ${ }^{4}$, Abhay K. Ram ${ }^{5}$\\
${ }^{1}$ Department of Physics, William \& Mary, Williamsburg, VA23185\\
${ }^{2}$ Department of Mathematics and Physical Sciences, Rogers State University, Claremore,OK 74017\\
${ }^{3}$ School of Electrical and Computer Engineering, National Technical University of Athens, Zographou 15780, Greece \\
${ }^{4}$ Department of Electrical \& Computer Engineering, Old Dominion University, Norfolk, VA 23529\\
${ }^{5}$ Plasma Science and Fusion Center, MIT, Cambridge, MA 02139\\}
\date{}

\begin{document}



\maketitle

\begin{abstract} 
It is well known that Maxwell equations can be expressed in a unitary Schrodinger-Dirac representation
for homogeneous media.  However, difficulties arise when considering inhomogeneous media.  A Dyson map points to a unitary field qubit basis, but the standard qubit lattice algorithm of interleaved unitary collision-stream operators must be augmented by some sparse non-unitary potential operators that recover the derivatives on the refractive indices.  The effect of the steepness of these derivatives on two dimensional scattering is examined with simulations showing quite complex wavefronts emitted due to transmissions/reflections within the dielectric objects.  Maxwell equations are extended to handle dissipation using Kraus operators.  Then, our theoretical algorithms are extended to these open quantum systems.  A quantum circuit diagram is presented as well as estimates on the required number of quantum gates for implementation on a quantum computer. 

\end{abstract}


\section{Introduction} 

Qubit lattice algorithms (QLA) were first being developed in the late 1990's to solve the Schrodinger equation \textbf{[1-3]} using unitary collision and streaming operators acting on some qubit basis.  QLA recovers the Schrodinger equation in the continuum limit to second order in the spatial lattice grid spacing.  Because the lattice node qubits are entangled by the unitary collision operator (much like in the formation of Bell states), QLA is encodable onto a quantum computer with an expected exponential speed-up over a classical algorithm run on a supercomputer.  Moreover, since QLA is extremely parallelizeable on a classical supercomputer, it provides an alternate algorithm for solving difficult problems in computational classical physics. 

We then applied these QLA ideas to the study of the nonlinear Schrodinger equation (NLS) \textbf{[4]}, by incorporating the cubic nonlinearity in the wave function, $|\psi|^2 \psi$, as an external potential operator following the unitary collide-stream operator sequence on the qubits. While the inclusion of such nonlinear terms poses no problem for a hybrid classical-quantum computer, it remains a very important and difficult research topic for their implementation on a quantum computer.  The accuracy of the QLA for NLS was tested for soliton-soliton collisions in long-term integration and compared to exact analytic solutions, and while the QLA is second order, it seemed to behave like a symplectic integrator.  The QLA was then extended to the totally integrable vector Manakov solitons 
\textbf{[5]} to handle inelastic soliton scattering.  The Manakov solitons are solutions to a coupled set of NLS equations.\\

Following these successful benchmarking simulations, we moved into QLA for two (2D) and three (3D) dimensional NLS equations  - where now there are no exact solutions to these nonlinear equations.  In the field of condensed matter, these higher dimension NLS equations are known as the Gross-Pitaevskii equations and give the mean field representation of the ground state wave function $\psi$ of a zero-temperature Bose-Einstein condensate (BEC).  For scalar quantum turbulence in 3D we \textbf{[6]} observed a triple energy cascade on a $5732^3$ grid, with the low-k ("classical") regime exhibiting a Kolmogorov $k^{-5/3}$ cascade in the \textit{compressible} kinetic energy while the incompressible kinetic energy exhibited a long k-range of $k^{-3}$ spectrum. Similar results were found for both 2D and 3D scalar quantum \textbf{[7-9]}, while results for spinor BECs can be found in \textbf{[10-12]}.  A somewhat related, but significantly different, approach is that of the quantum lattice Boltzmann method \textbf{[13-14]}.

Here we will discuss a QLA for the solution of Maxwell equations in a tensor dielectric medium \textbf{[15-18]}, and present some simulations results of the scattering of a 1D electromagnetic pulse off 2D localized  dielectric objects.  This can be viewed as a precursor to examining the scattering of electromagnetic pulses off plasma blobs in the exterior region of a tokamak.

 There has been much interest in rewriting the Maxwell equations in operator form and exploit its similarity to the Schrodinger-Dirac equation from the early 1930's (e.g., see the references in \textbf{[19]}).   For homogeneous media the qubit representation of the electric and magnetic fields, \textbf{E, H}, leads to a Dirac equation in a fully unitary representation.  
 However, when the media becomes 
inhomogeneous, a Dyson map \textbf{[20]} is required to yield a unitary Schrodinger-Dirac equation for the evolution of the electromagnetic qubit field representation.  In particular, one can use the fields
$(n_x E_x, n_y E_y, n_z E_z, B_x, B_y, B_z)$, where $n_i$ is the refractive index in the $i^{th}$-direction.

A QLA is developed for this representation of the Maxwell equations in Sec. 3.  This particular algorithm is a generalization of that used for the NLS equations.  The initial value problem is then solved for the case of an electromagnetic pulse propagating in the $x$-direction and scattering from different 2D localized dielectric objects with refractive index $n(x,y)$ in Sec. 4.  In particular, we have examined both polarizations of the pulse and $\nabla \cdot \mathbf{B} = 0$.  In Sec. 5 we consider the case in which the medium is dissipative.  This brings in the field of open quantum systems and interactions with an environment.  For illustration we consider a simplified cold electron-ion dissipative fluid model in an electromagnetic field.  Kraus operators are determined by a multidimensional analog of the quantum amplitude damping channel.  Some estimates on
the quantum gates required are given as well as a quantum circuit diagram illustrating the implementation of Kraus operators on the open Schrodinger equation.
We summarize our results in Sec. 6.

Finally, in this introduction, we quickly review the entanglement of qubits - and in particular for the 2-qubit Bell state \textbf{[21]}
\begin{equation}
B_{+}=\frac{1}{\sqrt2} (|00\rangle+|11\rangle)  .
\end{equation}
The most general 1-qubit states are $\{a_{0} |0\rangle + a_{1}  |1\rangle\}$ , and $\{b_{0} |0\rangle + b_{1}  |1\rangle\}$  with normalization 
$|a_0|^2 + |a_1|^2 = 1 = |b_0|^2 + |b_1|^2$.  The tensor product of these two 1-qubits yields a space of the form
\begin{equation}
a_{0} b_{0}  |00\rangle + a_{0} b_{1}  |01\rangle + a_{1} b_{0} |10\rangle+a_{1} b_{1} |11\rangle . 
\end{equation}
However the Bell state, Eq. (1) is not part of this tensor product space:  to remove the $ |01 \rangle$ state from Eq. (2) either $a_0 = 0$ or
$b_1 = 0$.  This in turn would remove either the $ |00 \rangle$ or the $ |11 \rangle$ states, respectively.  Now consider the unitary collision operator
\begin{equation}
C=\left[\begin{array}{cc}
\cos \theta & \sin \theta \\
-\sin \theta & \cos \theta
\end{array}\right]
\end{equation}
acting on the subspace basis $\{|00\rangle,|11\rangle\}$.  The choice of $\theta = \pi/4$ yields the Bell state $B_+$ - a maximally entangled state.  It is the quantum entanglement of states that will give rise to exponential speed-up of a quantum algorithm.  The QLA is a sequence of interleaved unitary collision-streaming operators that entangle the qubits and then spread that entanglement throughout the lattice.

\section{The Dyson map and the generation of a unitary evolution equation for Maxwell equations}

Consider the subset of Maxwell equations 
\begin{equation}
\nabla \times \mathbf{E} = - \frac{\partial \mathbf{B}}{\partial t}   \quad , \quad   \nabla \times \mathbf{H} =  \frac{\partial \mathbf{D}}{\partial t} 
\end{equation}
and treat $\nabla \cdot \mathbf{B} = 0$ and $\nabla \cdot \mathbf{D} = 0 $ as initial constraints that remain satisfied in the continuum limit for 
all times.  (This, of course,  follows immediately from taking the divergence of Eq. (4) ).

For lossless media, the electric and magnetic fields satisfy the constitutive relations for a tensor dielectric non-magnetic medium
\begin{equation}
\mathbf{D}=\boldsymbol{\varepsilon} \cdot \mathbf{E}, \quad \mathbf{B}=\mu_{0} \mathbf{H} .
\end{equation}
For Hermitian  $\boldsymbol{\varepsilon}$ one can transform to a  coordinate system in which  $\boldsymbol{\varepsilon}$ is diagonal.  
Equation (5) can be 
rewritten in matrix form
\begin{equation}
\mathbf{d}=\mathbf{W \, u} \quad , \text{with} \quad \mathbf{d} \doteq (\mathbf{D}, \mathbf{B})^{\mathbf{T}} , \quad
\mathbf{u} \doteq (\mathbf{E}, \mathbf{H})^{\mathbf{T}}
\end{equation}
where \textbf{W} is a $6 \times 6$ Hermitian block diagonal constitutive matrix
\begin{equation}
\mathbf{W}=\left[\begin{array}{cc}
\boldsymbol{\varepsilon} \mathbb{I}_{3 \times 3} & 0_{3 \times 3} \\
0_{3 \times 3}& \mu_{0} \mathbb{I}_{3 \times 3}
\end{array}\right]  .
\end{equation}
$\mathbb{I}_{3 \times 3} $ is the $3 \times 3$ identity matrix, and the superscript $\mathbf{T}$ in Eq. (6) is the transpose.  In matrix
form, the Maxwell equations, Eq. (4) become
\begin{equation}
i \frac{\partial \mathbf{d}}{\partial t}=\mathbf{M \, u}
\end{equation}
\noindent where, under standard boundary conditions, the curl-matrix operator $\mathbf{M}$ is Hermitian :
\begin{equation}  
\mathbf{M}=\left[\begin{array}{cc}
0_{3 \times 3} & i \nabla \times \\
-i \nabla \times & 0_{3 \times 3}
\end{array}\right]  .
\end{equation}
From Eq. (6), since $\mathbf{W^{-1}}$ exists, $\mathbf{u} = \mathbf{W}^{-1}  \mathbf{d} $, so that Eq. (8) can be written
\begin{equation}
i \frac{\partial \mathbf{u}}{\partial t}=\mathbf{W}^{-\mathbf{1}} \mathbf{M} \mathbf{u}
\end{equation}
If the medium is homogeneous, then $\mathbf{W}^{-1}$ is constant and will commute with the curl-operator $\mathbf{M}$.
Under these conditions, the product $\mathbf{W}^{-1} \mathbf{M}$ is Hermitian and Eq. (10) gives unitary evolution for $\mathbf{u} = 
(\mathbf{E, H})^T$.\\

However, if the medium is spatially inhomogeneous, then $\left[\mathbf{W}^{-1},\mathbf{M} \right] \neq 0$
and the evolution equation for the $\mathbf{u}$-field is not unitary.

\subsection{Dyson map, [20]}
To determine a unitary evolution of the electromagnetic fields in an inhomogeneous dielectric medium, it \textbf{[20]} has been shown
that there exists a Dyson map $\boldsymbol{\rho}$: $\mathbf{u} \rightarrow \mathbf{Q}$ such that in the new field variables $\mathbf{Q}$
the resulting evolution equation will be unitary.  For the Maxwell equations consider
\begin{equation}
\mathbf{Q} = \boldsymbol{\rho} \mathbf{u}  = \mathbf{W}^{1/2} \mathbf{u}.
\end{equation}
For time-independent media, the evolution equation for the new fields $\mathbf{Q}$  is 
\begin{equation}
i \boldsymbol{\rho}\frac{\partial \mathbf{u}}{\partial t}=\boldsymbol{\rho} \mathbf{W}^{-\mathbf{1}} \mathbf{M} \boldsymbol{\rho}^{-1} \boldsymbol{\rho}\mathbf{u} \quad \Rightarrow 
 i \frac{\partial \mathbf{Q}}{\partial t}=\boldsymbol{\rho} \mathbf{W}^{-\mathbf{1}} \mathbf{M} \boldsymbol{\rho}^{-1} \mathbf{Q}
\end{equation}
and is indeed unitary.  Explicitly, the new fields, Eq. (11) and the $\boldsymbol{\rho}$  are
\begin{equation}
\mathbf{Q} = \left[\begin{array}{cccccc}
q_0 \\
q_1\\
q_2 \\
q_3  \\
q_4 \\
q_5 
\end{array}\right] = \left[\begin{array}{cccccc}
n_x E_x \\
n_y E_y \\
n_z E_z \\
\mu_0^{1/2} H_x  \\
\mu_0^{1/2} H_y  \\
\mu_0^{1/2} H_z  
\end{array}\right] \quad , \quad 
\boldsymbol{\rho} = \left[\begin{array}{cccccc}
n_x & 0 & 0 & 0& 0& 0 \\
0 & n_y & 0 & 0 & 0 & 0 \\
0 & 0 & n_z & 0 & 0 & 0 \\
0 & 0 & 0 & \mu_0^{1/2}  &  0 & 0  \\
0 & 0 & 0 & 0 & \mu_0^{1/2} & 0 \\
0 & 0 & 0 & 0 & 0 & \mu_0^{1/2} 
\end{array}\right] 
\end{equation}
The refractive index  $n_i =  \sqrt{ \varepsilon_i}$.  Typically we will use units where $\mu_0 = 1$.  In component form, Maxwell equations for
fields and with constitutive matrix restricted to spatially 2D $(x,y)$ dependence, Eq. (12) reduces to
\begin{equation}  
\begin{aligned}
\frac{\partial q_0}{\partial t} = \frac{1}{n_x} \frac{\partial q_5}{\partial y} , \qquad
\frac{\partial q_1}{\partial t} = - \frac{1}{n_y} \frac{\partial q_5}{\partial x} , \qquad
\frac{\partial q_2}{\partial t} =  \frac{1}{n_z} \left[ \frac{\partial q_4}{\partial x} -\frac{\partial q_3}{\partial y} \right] \\
\frac{\partial q_3}{\partial t} = - \frac{\partial (q_2/n_z)}{\partial y} , \qquad
\frac{\partial q_4}{\partial t} = \frac{\partial (q_2/n_z)}{\partial x} , \qquad
\frac{\partial q_5}{\partial t} = - \frac{\partial (q_1/n_y)}{\partial x}  + \frac{\partial (q_0/n_x)}{\partial y} 
\end{aligned}
\end{equation}

\section{A Qubit Lattice Representation for 2D Tensor Dielectric Media}   

QLA consists of a sequence of unitary collision and streaming operators on a 2D spatial lattice which will recover the continuum Maxwell equations, Eq. (14) to second order in the spatial grid size, $\delta$. In particular we need to have 6 qubits/lattice site to represent the fields components in Eq. (13).
QLA permits us to handle the x- and y-dependence separately.  Let us first consider the x-dependence  and recover the $\partial q_i/\partial x$ - terms.  From Eq. (14) we see coupling between $q_1 \leftrightarrow q_5$, $q_2\leftrightarrow q_4$, Hence we introduce the local entangling collision operator
\begin{equation}
C_X=\left[\begin{array}{cccccc}
1 & 0 & 0 & 0& 0& 0 \\
0 & cos \,\theta_1 & 0 & 0 & 0 & - sin\,\theta_1 \\
0 & 0 & cos\,  \theta_2 & 0 & - sin \,\theta_2 & 0 \\
0 & 0 & 0 & 1 &  0 & 0  \\
0 & 0 & sin\,\theta_2 & 0 & cos\, \theta_2 & 0 \\
0 & sin\, \theta_1 & 0 & 0 & 0 & cos \,\theta_1
\end{array}\right]
\end{equation}
The collision angles $\theta_1$ and $\theta_2$ need to be chosen to recover the refractive index factors before the corresponding spatial derivatives,
\begin{equation}
\theta_1 = \frac{\delta}{4n_y} \quad , \quad  \theta_2 = \frac{\delta}{4n_z}  .
\end{equation}
The first of the unitary streaming operators will stream qubits $q_1, q_4$ one lattice unit in either direction
while leaving the other four qubits fixed : $S_{14}^{\pm}$.  The other unitary streaming operator will act on qubits $q_2, q_5$ : $S_{25}^{\pm}$.  
The final unitary collide-stream sequence, $\mathbf{U_X}$ in the x-direction that leads to a second order scheme in $\delta$ can be shown to be
\begin{equation}
\mathbf{U_X} = S^{+x}_{25}.C_X^\dag . S^{-x}_{25}.C_X. S^{-x}_{14}.C_X^\dag . S^{+x}_{14}.C_X .S^{-x}_{25}.C_X . S^{+x}_{25}.C_X^\dag. S^{+x}_{14}.C_X . S^{-x}_{14}.C_X^\dag .
\end{equation}
It should be noted that if only applies the first 4 collide-stream sequence in Eq. (17) then the algorithm would only be first order accurate.

Similarly, to recover the $\partial q_i/\partial y$-terms one would collisionally entangle qubits $q_0 \leftrightarrow q_5$, $q_2\leftrightarrow q_3$ 
 with
 \begin{equation}   
C_Y=\left[\begin{array}{cccccc}
cos\, \theta_0 & 0 & 0 & 0& 0& sin\, \theta_0 \\
0 & 1 & 0 & 0 & 0 & 0 \\
0 & 0 & cos\,  \theta_2 &sin \,\theta_2 & 0 &   0 \\
0 & 0 & -sin\,\theta_2 & cos \,\theta_2 &  0 & 0  \\
0 & 0 & 0& 0 & 1 & 0 \\
-sin \,\theta_0& 0 & 0 & 0 & 0 & cos \,\theta_0
\end{array}\right]   ,
\end{equation}
with corresponding collision angles $\theta_0$ and $\theta_2$.  $\theta_2$ is given in Eq. (16), and
\begin{equation}
\theta_0 = \frac{\delta}{4 n_x}  .
\end{equation}
The streaming operator $S^{y}_{03}$ will act on qubits $q_0, q_3$ only and similarly for the operator $S^{y}_{25}$.  The final unitary collide-stream
second-order accurate or the y-direction for Maxwell equations is
\begin{equation}
\mathbf{U_Y} = S^{+y}_{25}.C_Y^\dag . S^{-y}_{25}.C_Y. S^{-y}_{03}.C_Y^\dag . S^{+y}_{03}.C_Y .S^{-y}_{25}.C_Y . S^{+y}_{25}.C_Y^\dag. S^{+y}_{03}.C_Y . S^{-y}_{03}.C_Y^\dag 
\end{equation}

	We still need to recover the spatial derivatives on the refractive index components in Eq. (14).  To obtain the $\partial n_z/\partial x$ and $\partial n_y/\partial x$ 
terms we introduce the (non-unitary) sparse potential matrix
\begin{equation}
V_X=\left[\begin{array}{cccccc}
1 & 0 & 0 & 0& 0& 0 \\
0 & 1 & 0 & 0 & 0 & 0\\
0 & 0 & 1 & 0 &0& 0 \\
0 & 0 & 0 & 1 &  0 & 0  \\
0 & 0 &- sin\,\beta_2 & 0 & cos\, \beta_2 & 0 \\
0 & sin\, \beta_0 & 0 & 0 & 0 & cos \,\beta_0
\end{array}\right]
\end{equation}
with collision angles
\begin{equation}   
\beta_0 = \delta^2 \frac{\partial n_y/\partial x}{n^2_y} \quad , \ \beta_2 = \delta^2 \frac{\partial n_z/\partial x}{n^2_z} ,
\end{equation}
while the corresponding (non-unitary) sparse potential matrix to recover the $\partial/\partial y$-derivatives in the refractive index components is
\begin{equation}
V_Y=\left[\begin{array}{cccccc}
1 & 0 & 0 & 0& 0& o \\
0 & 1 & 0 & 0 & 0 & 0\\
0 & 0 & 1 & 0 &0& 0 \\
0 & 0 & \cos\, \beta_3& \sin \, \beta_3 &  0 & 0  \\
0 & 0 & 0 & 0 & 1 & 0 \\
-sin\, \beta_1 & 0 & 0 & 0 & 0 & cos \,\beta_1
\end{array}\right]
\end{equation}
with collision angles
\begin{equation}  
 \beta_1 = \delta^2 \frac{\partial n_x/\partial y}{n^2_x} \quad , \quad \beta_3 = \delta^2 \frac{\partial n_z/\partial y}{n^2_z}  .
\end{equation}	
	
	Thus the final discrete QLA,  that models the 2D Maxwell equations, Eq. (14), to $O(\delta^2)$, advances the lattice qubit-vector $\mathbf{Q}(t)$ to 
$\mathbf{Q}(t+\Delta t)$ is
\begin{equation}
\mathbf{Q}(t+\Delta t)  = V_Y . V_X. \mathbf{U_Y} . \mathbf{U_X} . \mathbf{Q}(t)
\end{equation}
provided we have diffusion ordering in the space-time lattice:  i.e., $\Delta t = \delta^2$.	It is this ordering that requires us to have the unitary collision angles
to be $O(\delta)$, Eqs. (16) and (19), and the external potential angles $O(\delta^2)$, Eqs. (22).  We note that computationally QLA is more accurate if we employ the external potentials twice:  once half-way through the collide-stream sequence and then at the end.

\subsection{Nonunitary External Potential Operators, Eqs. (21) and (23).}    
Recently, considerable effort has been expended into developing more efficient approximation to handling the evolution operator of a complex Hamiltonian 
system than the standard Suzuki-Trotter expansion \textbf{[22]}.  In particular the idea \textbf{[23, 24]} has been floated of approximating the full unitary opertor by a \textit{sum} of unitary operators.  The actual implementation onto a quantum computer we will leave to another paper, as one of the outcomes of QLA discussed here will be a quantum-inspired highly efficient classical supercomputer algorithm.  Moreover, its encoding onto a quantum computer will require error-correcting qubits with long coherence times, something currently out of reach in the noisy qubit regime we are in..  Here we will show the 4 unitary operators needed whose sum yields the sparse non-unitary potential operator $V_X$, Eq. (21).  Letting $\mathbb{I}_6$ be the $6 \times 6$ identity matrix, then it is easily verified that
\begin{equation}
V_X = \frac{1}{2} \sum_{i=1}^4 LCU_i
\end{equation}
where the first two unitaries are diagonal
\begin{equation}
LCU_1 = \mathbb{I}_6 \quad , \quad LCU_2= diag \left(-1, 1,1, -1, -1, -1  \right)
\end{equation}
and the remaining two unitaries are
\begin{equation}
LCU_3=\left[\begin{array}{cccccc}
1 & 0 & 0 & 0& 0& o \\
0 & \cos \beta_0 & 0 & 0 & 0 & - \sin \beta_0\\
0 & 0 & \cos \beta_2 & 0 &\sin \beta_2 & 0 \\
 0 & 0 & 0& 1 & 0 & 0 \\
0 & 0 & -\sin \beta_2 & 0 &  \cos \beta_2 & 0  \\
0 & \sin \beta_0 & 0 & 0 & 0 & \cos \beta_0
\end{array}\right] ,  
\end{equation}
and
\begin{equation}
LCU_4=\left[\begin{array}{cccccc}
1 & 0 & 0 & 0& 0& o \\
0 &- \cos \beta_0 & 0 & 0 & 0 &  \sin \beta_0\\
0 & 0 & -\cos \beta_2 & 0 &-\sin \beta_2 & 0 \\
 0 & 0 & 0& 1 & 0 & 0 \\
0 & 0 & -\sin \beta_2 & 0 &  \cos \beta_2 & 0  \\
0 & \sin \beta_0 & 0 & 0 & 0 & \cos \beta_0
\end{array}\right] .
\end{equation}

\subsection {Conservation of Energy}
In a fully unitary representation, the norm of \textbf{Q} is a constant of the motion.   This is simply the conservation of energy of the electromagnetic field.  For fields being a function of $(x,y)$, we have from Eqs. (11)-(13)
\begin{equation}  
\mathcal{E}(t) =\frac{1}{L^2} \int_0^L dxdy \mathbf{Q} \cdot \mathbf{Q}
= \frac{1}{L^2} \int_0^L dx dy \left[\epsilon_x E_x^2 +\epsilon_y E_y^2 + \epsilon_z E_z^2 + \mu_o \left(B_x^2+B_y^2+B_z^2 \right) \right]
\end{equation}
where the (diagonal) tensor dielectric $\epsilon_x = n_x^2, ...$ and we restrict ourselves to non-magnetic materials, for simplicity.

\section{2D Numerical Simulations from the QLA for Electromagnetic Scattering from 2D dielectric objects}

We now present detailed QLA simulations to the initial value problem of the scattering of a 1D electromagnetic pulse from a localized dielectric object.
In particular we consider a 1D Gaussian pulse propagating in the $x$-direction towards a localized dielectric object of refractive index $n(x,y)$.   The initial pulse has non-zero field components $E_z, B_y$,, Fig. 1,  and scatters from either a localized cylindrical dielectric, Fig. 2a, or a conic dielectric object, Fig 2b.  These simulations were performed for $\delta = 0.1$.

\begin{figure}[!b]\centering
\includegraphics[width=4.4in]{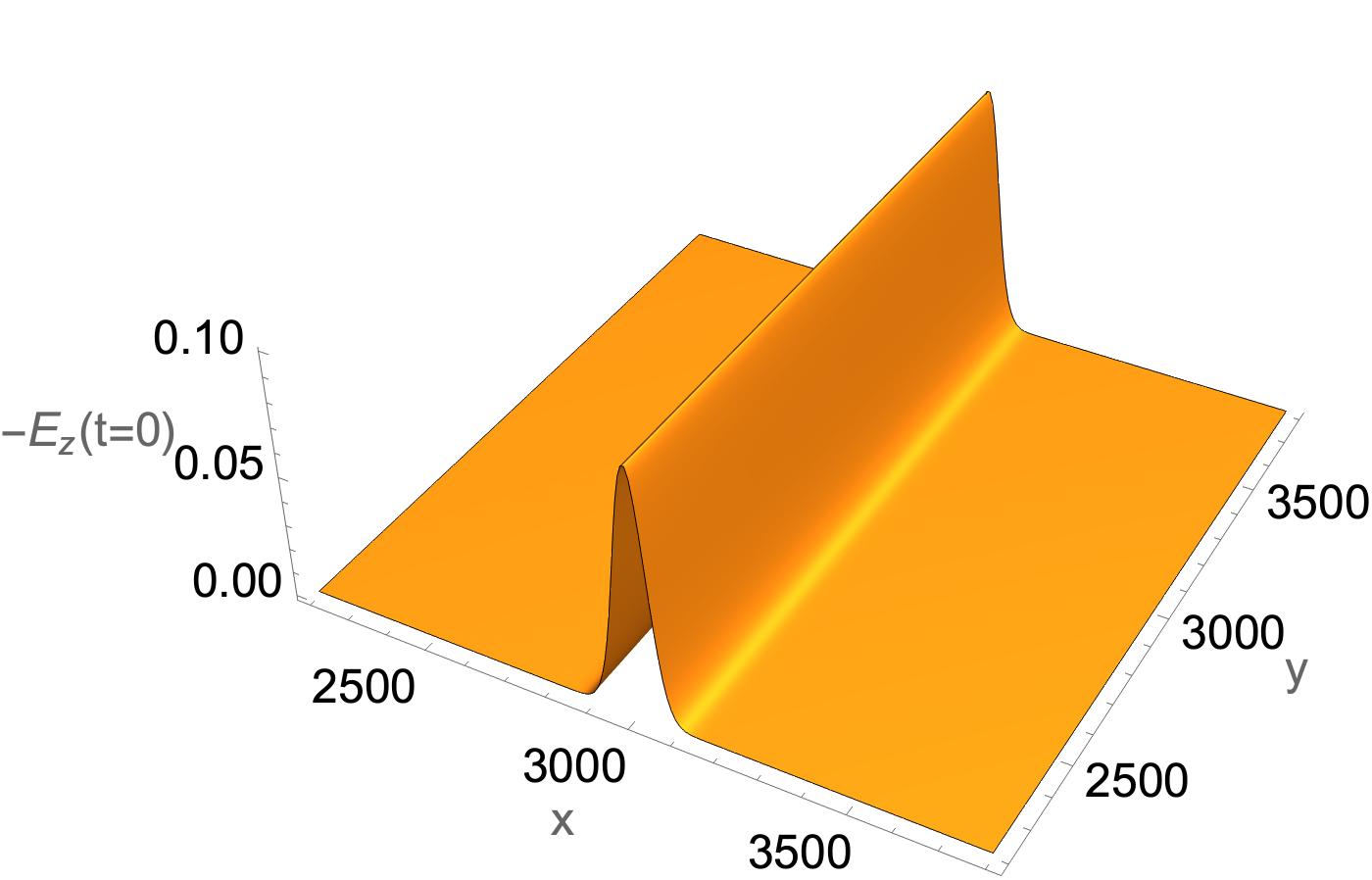}
\caption{ A 1D electromagnetic pulse with initial fields $-E_z(x,t=0), B_y(x,t=0)$ .  2D Simulation grid $L \times L$ with $L = 8192$.  Pulse full-width (in lattice units) $\approx 200$.  Since the Maxwell equations are linear and homogenous, the initial amplitude of the fields is arbitrary.
}
\end{figure}

\begin{figure}[!b]\centering
\includegraphics[width=3.2in]{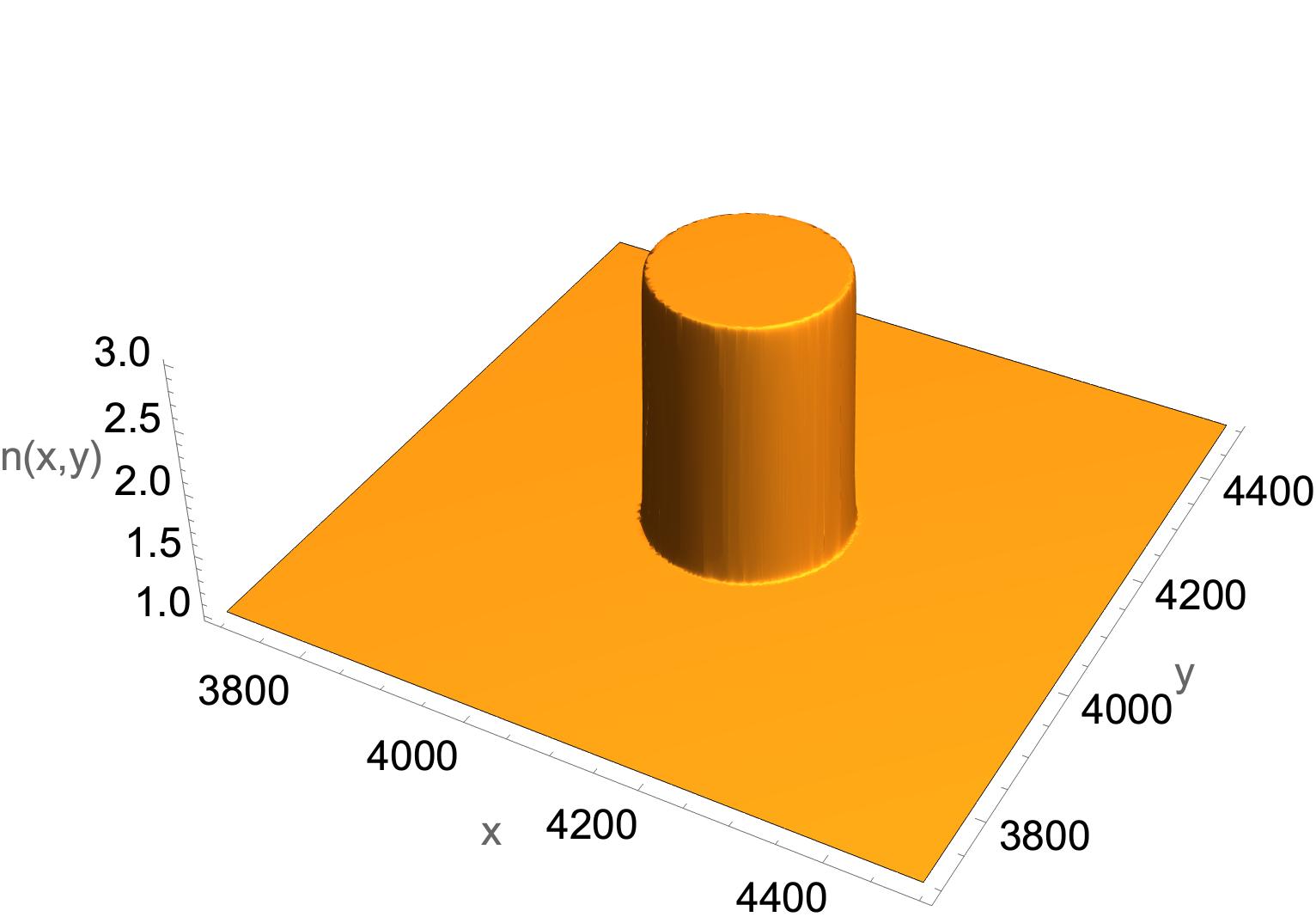}
\includegraphics[width=3.2in]{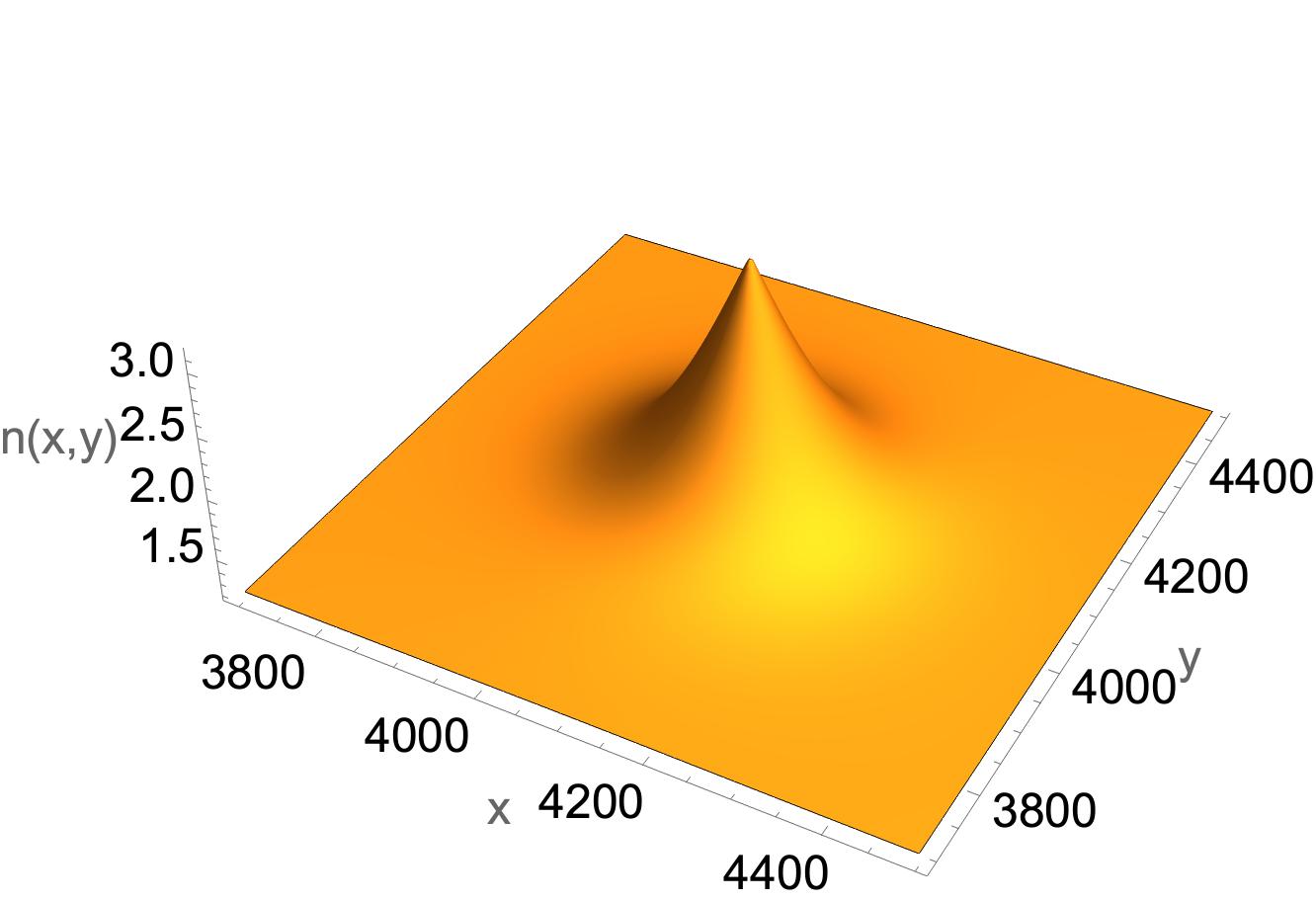}
(a)  localized dielectric cylinder  \qquad \qquad  \qquad  (b)  dielectric cone 
\caption{ (a) the dielectric cylinder, diameter $\approx 200$, has rapidly increasing boundary dielectric from vacuum $n=1$ to $n_{max} = 3$.
whereas (b) the conic dielectric, base $\approx 240$, has smoothly increasing dielectric from vacuum to conic peak of $n_{max} = 3$.
}
\end{figure}
It should be noted that QLA is an initial value algorithm :  the refractive index profiles are smooth (e.g., hyperbolic tangents for the dielectric cylinder with boundary layer thickness $\approx 10$ lattice units) and so \textit{no} internal boundary conditions are imposed at any time in the simulation.

\subsection{Effects of broken symmetry}
When the 1D pulse scatters off the dielectric object with refractive index $n(x,y)$ the initial electric field spatial dependence $E_z(x)$ now becomes a function $E_z(x,y)$, while the initial magnetic field $B_y(x)$ will become a function $B_y(x,y)$.  Now the scattered field
has $\partial B_y/\partial y \neq 0$.  Thus for $\nabla \cdot \mathbf{B} = 0$, the scattered field must develop an appropriate $B_x(x,y)$.  This is seen in our QLA simulations, even though the explicit discrete collide-stream algorithm only models asymptotically the Maxwell subset, Eq. (4).  $\nabla \cdot \mathbf{D} = 0$ and this is exactly conserved in our QLA simulation.

Similarly for initial $E_y(x)$ polarization.  In this case the scattered magnetic field $B_z(x,y)$ satisfies, for our 2D  scattering in the x-y plane, $\nabla \cdot \mathbf{B} = 0$ exactly and no other magnetic field components are generated.  Our discrete QLA recovers this 
$\nabla \cdot \mathbf{B} = 0$ exactly.  However, in an attempt to preserve $\nabla \cdot \mathbf{D} = 0$, the QLA will generate a non-zero $E_x(x,y)$ field.

\subsection{Scattering from localized 2D dielectric objects with refractive index $n(x,y)$}
Consider a 1D pulse with polarization $E_z(x,t)$ propagating in a vacuum towards a 2D dielectric scatterer.  In Figs. 3-6 we consider the time evolution of the resultant scattered $E_z$-field.  The initial pulse is followed for a short time while it is propagating in the vacuum to verify that the QLA correctly determines its motion.  As part of the pulse interacts with the dielectric object, the pulse speed within the dielectric itself is decreased by the inverse of the refractive index profiles, $n(x,y)$.  The remainder of the 1D pulse propagates undisturbed since it is still propagating within the vacuum.  

One sees in Fig. 3a a circular-like wavefront reflecting back into the vacuum, with its $E_z$ field $\pi$ out of phase as the reflection is occurring from a low to higher refractive index around the vacuum-cylinder interface.  One does not find such a reflected wavefront when the pulse interacts with the conic dielectric, Fig. 3b.

\begin{figure}[!b]\centering
\includegraphics[width=3.2in]{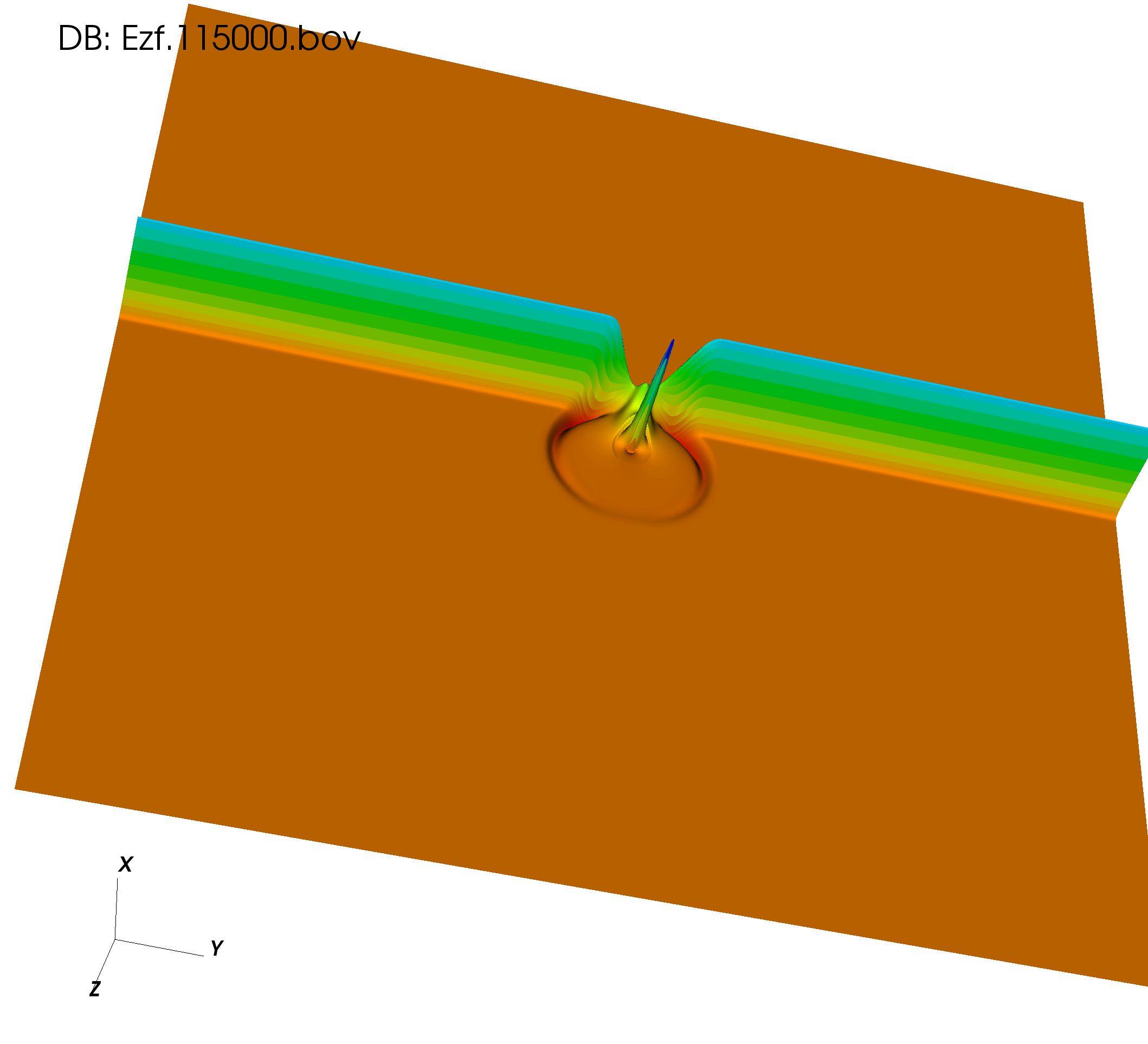}
\includegraphics[width=3.2in]{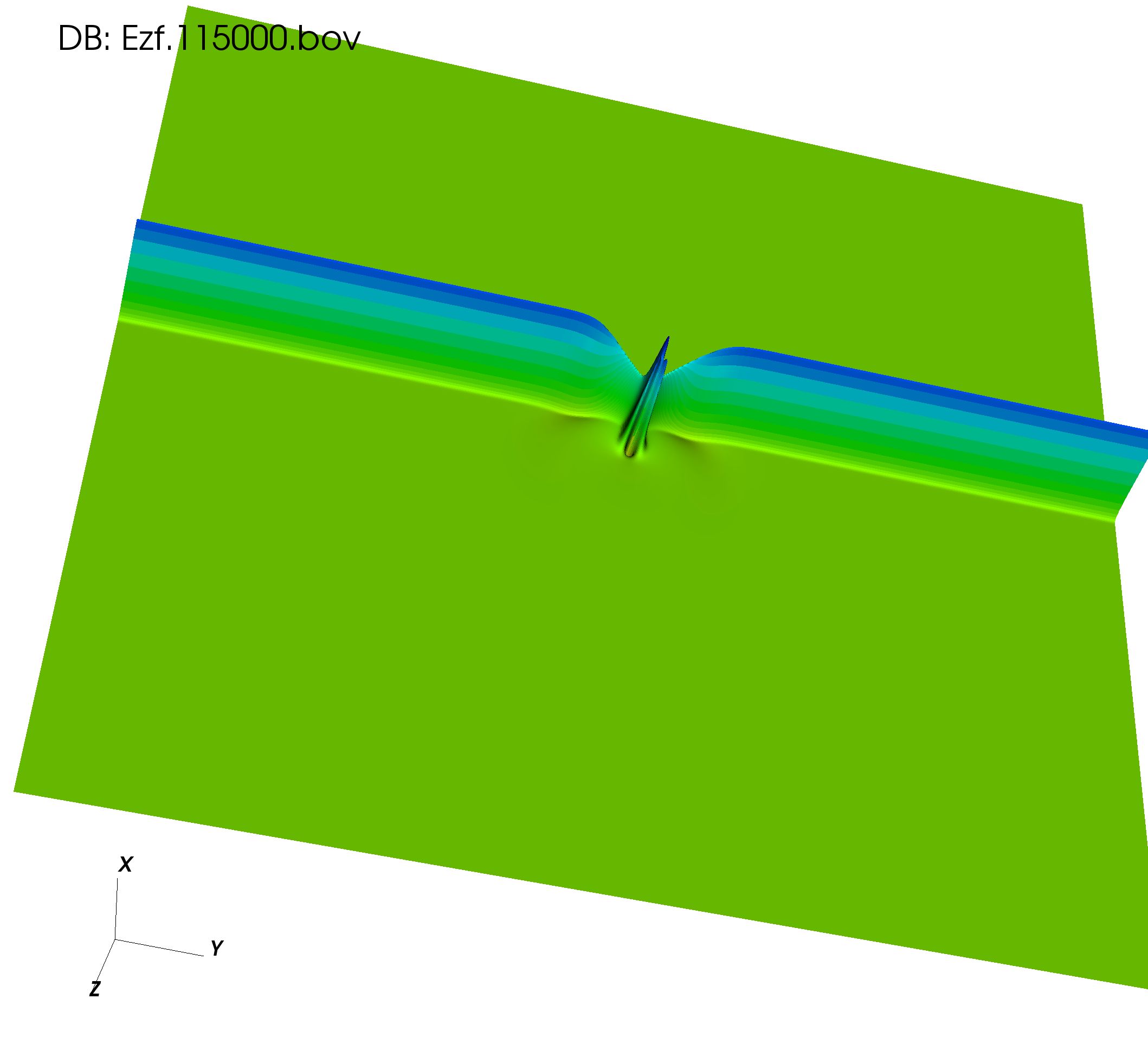}
(a)  dielectric cylinder  \qquad \qquad  \qquad  (b)  dielectric cone 
\caption{The scattered $E_z$ field after 15,000 iterations (i.e., $t=15k$).  (a)  There is an internal reflection at the back of the cylindrical dielectric.
}
\end{figure}

At $t = 23.4k$, there is a major wavefront emanating from the back of the cylindrical dielectric, Fig. 4a.  For the conic dielectric there is a major wavefront 
reflected from the apex of the conic dielectric, and this propagates out of the cone with little reflection, Fig. 4b.

\begin{figure}[!b]\centering
\includegraphics[width=3.2in]{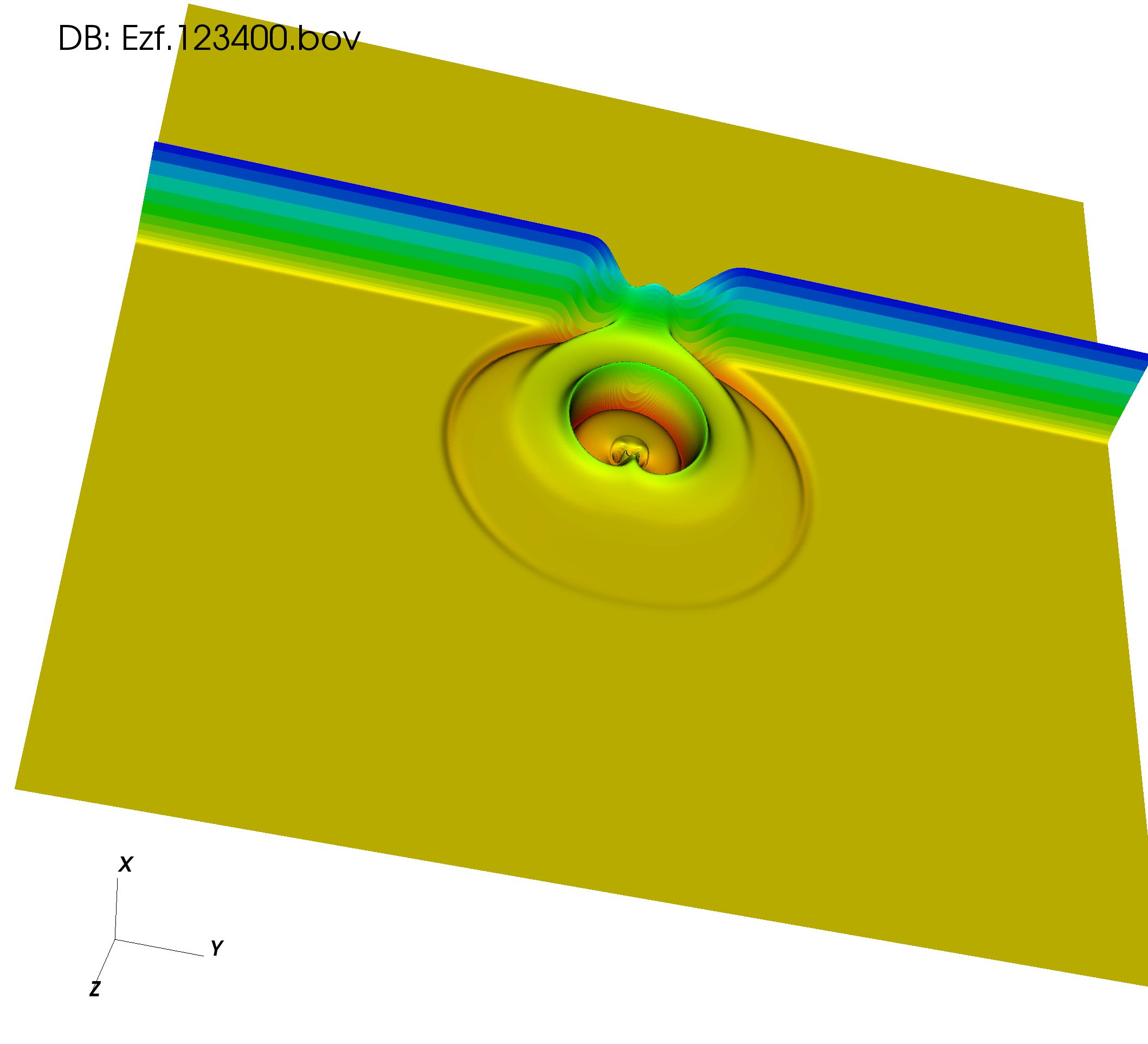}
\includegraphics[width=3.2in]{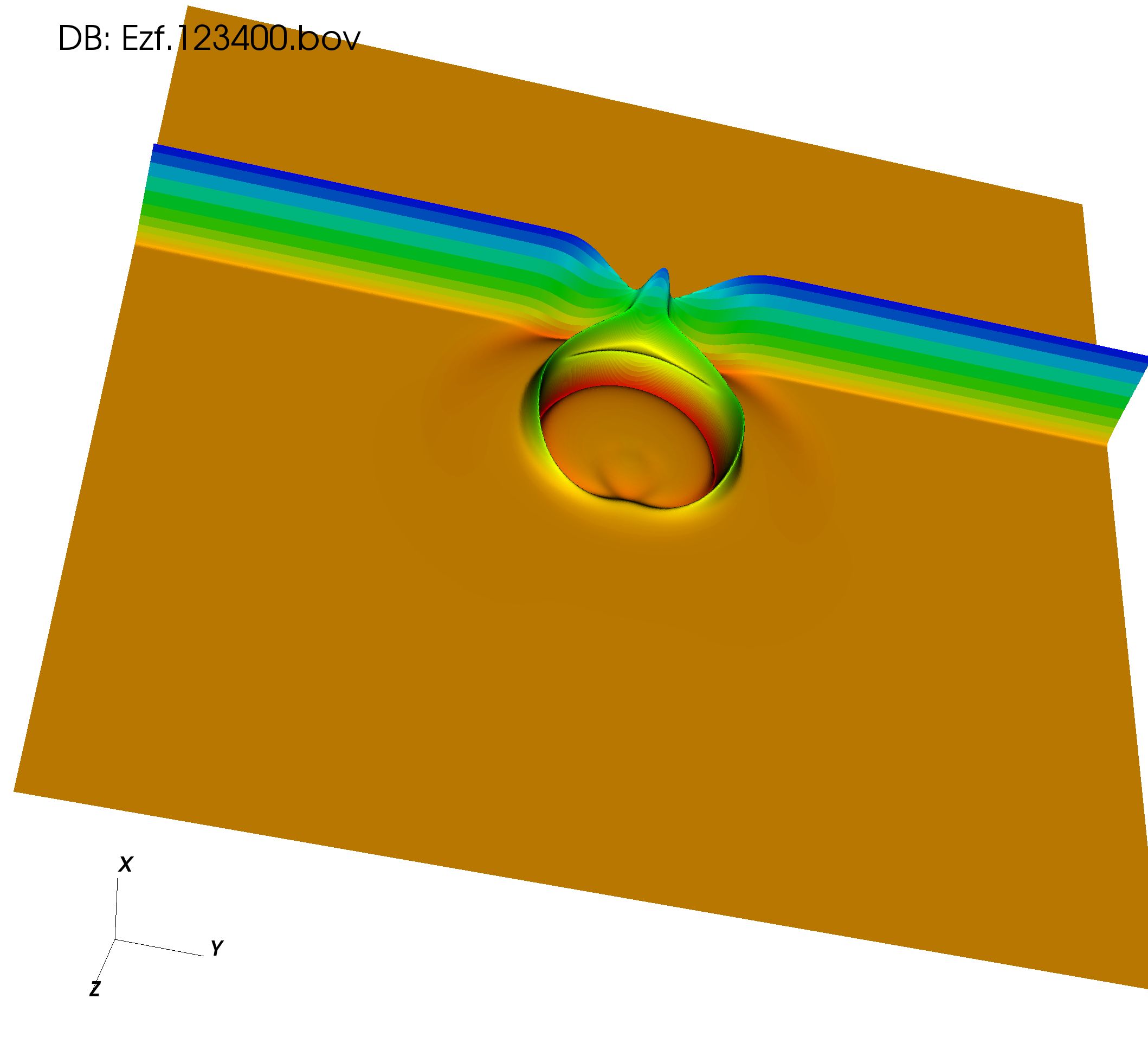}
(a)  dielectric cylinder  \qquad \qquad  \qquad  (b)  dielectric cone 
\caption{The scattered $E_z$ field at $t = 23.4k$..  (a)  A reflected circular wavefront occurs as that part of the pulse
reaches the back-end of the cylindrical dielectric, along with the initial reflected circular wavefront with its $\pi$-changed phase at the front of the vacuum-cylinder boundary.  (b)  For the conic dielectric, there is an internal reflection from the apex of the cone's $n_{max}$ which then propagates out of the weakly varying cone edges.
}
\end{figure}

One clearly sees at $t = 31.2k$ that more $E_z$ wavefronts are being created because of the large refractive index gradients at the vacuum-cylinder dielectric boundary, while such gradients are missing from the vacuum-cone interface which leads to no new wavefronts in the scattering off the dielectric cone, Fig.5. 
\begin{figure}[!b]\centering
\includegraphics[width=3.2in]{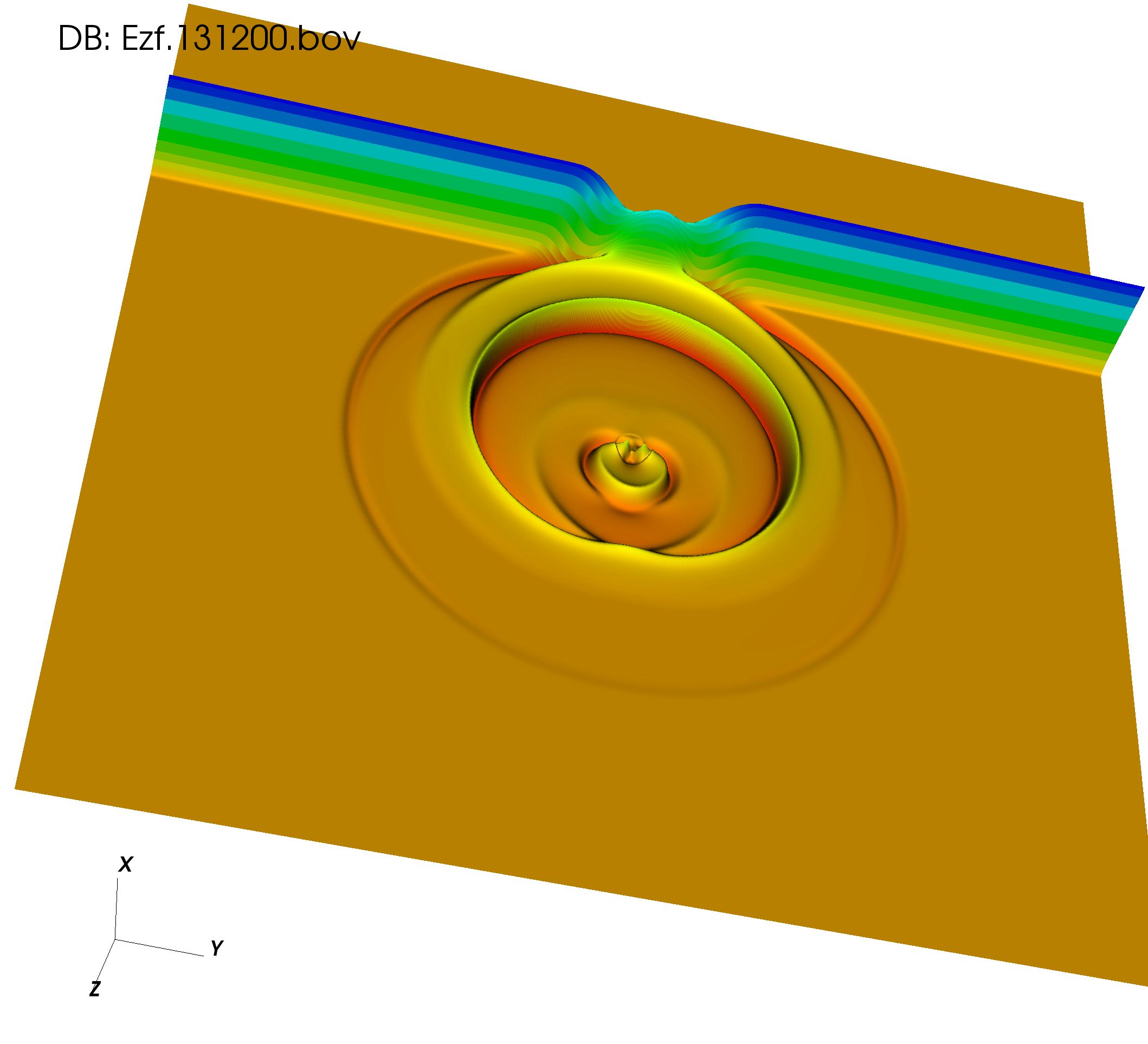}
\includegraphics[width=3.2in]{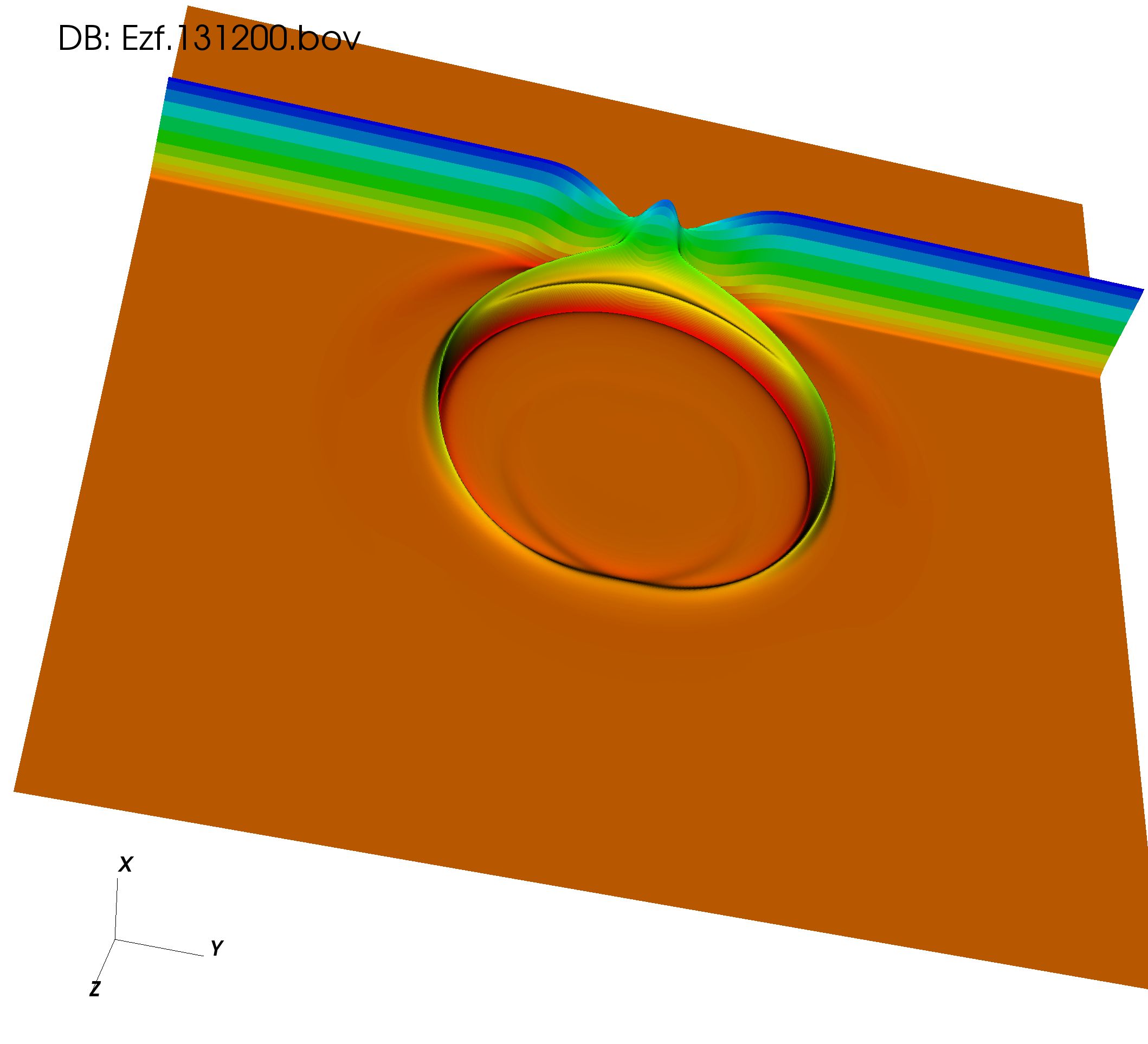}
(a)  dielectric cylinder  \qquad \qquad  \qquad  (b)  dielectric cone 
\caption{The scattered $E_z$ field at $t = 31.2k$.  (a)  There are multiple reflections/transmission within the boundaries of the dielectric cylinder. (b)  There is only one major reflection from the apex of the cone and which then propagates readily out from the cone edge's.
}
\end{figure}

At $t = 49.2k$, the complex $E_z$ wavefronts are due to repeated reflections and transmissions from the cylinder dielectric, Fig. 5a.  However, because of the slowly changing boundaries of the dielectric cone there are no more reflections and one sees only the outgoing
wavefront from the pulses interaction with the region around the $n_{max}$ of the cone.  Since the pulse reaches the apex of the cone
before the corresponding pulse hits the backend of the dielectric cylinder, the conic wavefront is further advanced than that of the cylindrical wavefront, Fig. 5b.

\begin{figure}[!b]\centering
\includegraphics[width=3.2in]{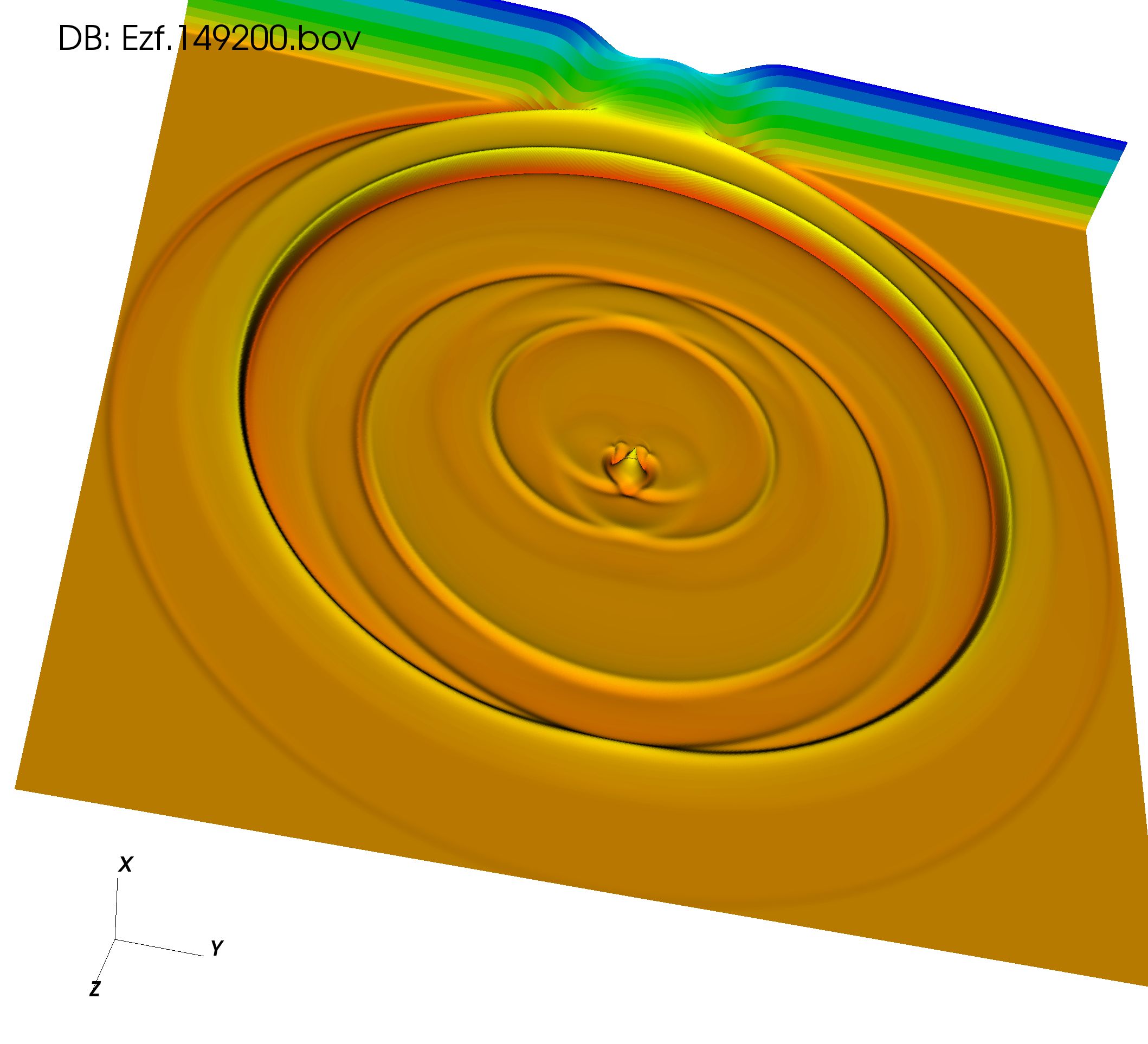}
\includegraphics[width=3.2in]{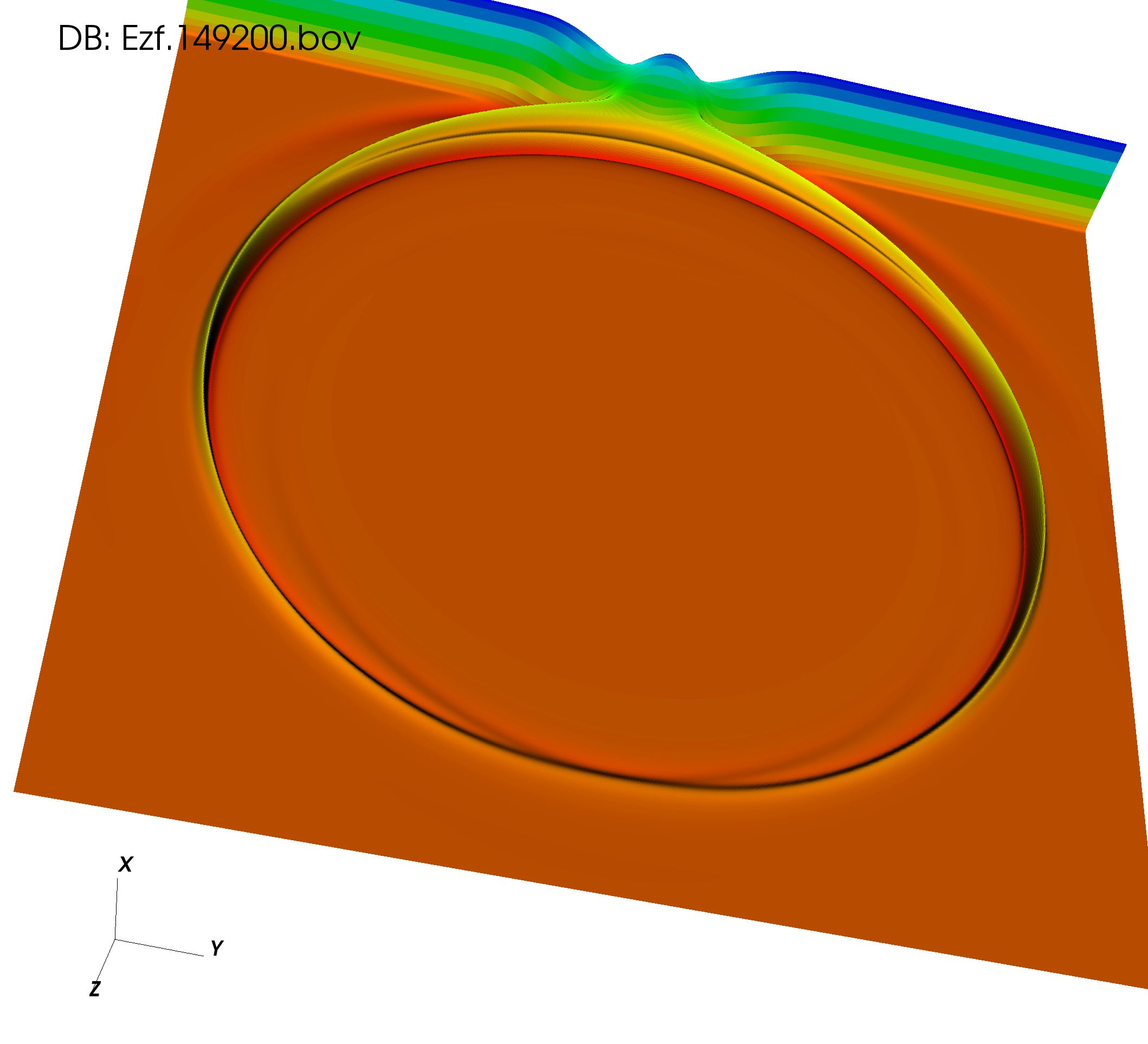}
(a)  dielectric cylinder  \qquad \qquad  \qquad  (b)  dielectric cone 
\caption{The scattered $E_z$ field at $t = 49.2k$.  (a)  There are multiple reflections/transmission within the boundaries of the dielectric cylinder. (b)  There is only one major reflection from the apex of the cone and which then propagates readily out from the cone edge's.
}
\end{figure}

\subsubsection{Auxiliary fields and $\nabla \cdot \mathbf{B}$}  
For incident $E_z$ polarization and with 2D refractive index $n(x,y)$, the scattered electromagnetic fields will need to generate a $B_x$ field in order to have $\nabla \cdot \mathbf{B} = 0$.  In Fig. 6a and 6b we plot the self-generated $B_x(x,y)$ field at $t = 23.4k$  and $t=49.2k$ for scattering from the dielectric cylinder.  It is also found that $| \nabla \cdot \mathbf{B} |/ B_0$ is typically zero everywhere in the spatial lattice except for a very localized region around the vacuum-dielectric boundary layer where the normalized $max(|\nabla \cdot \mathbf{B}|)/B_0$ reaches around 0.01 at very few isolated grid points.
\begin{figure}[!b]\centering
\includegraphics[width=3.2in]{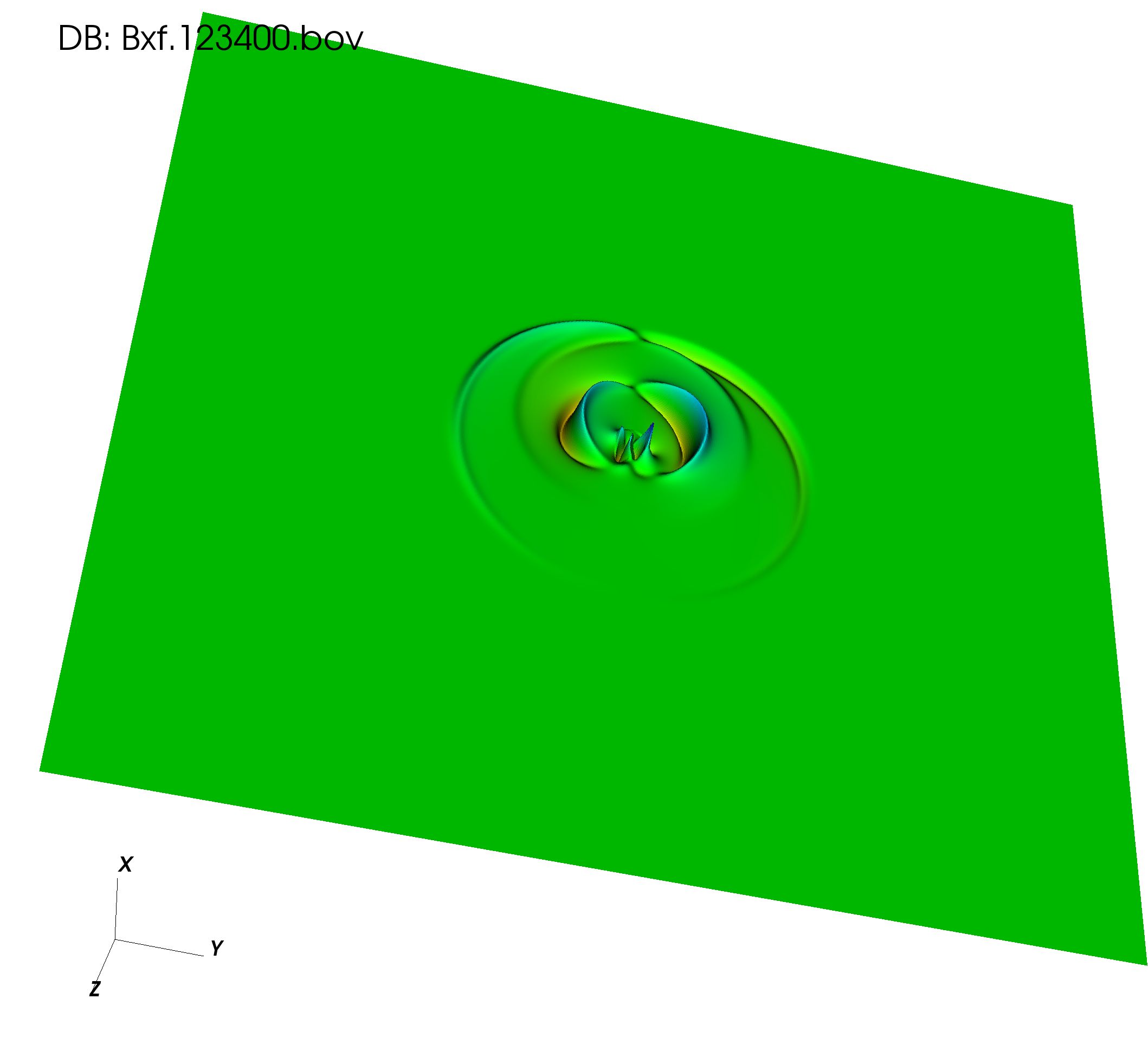}
\includegraphics[width=3.2in]{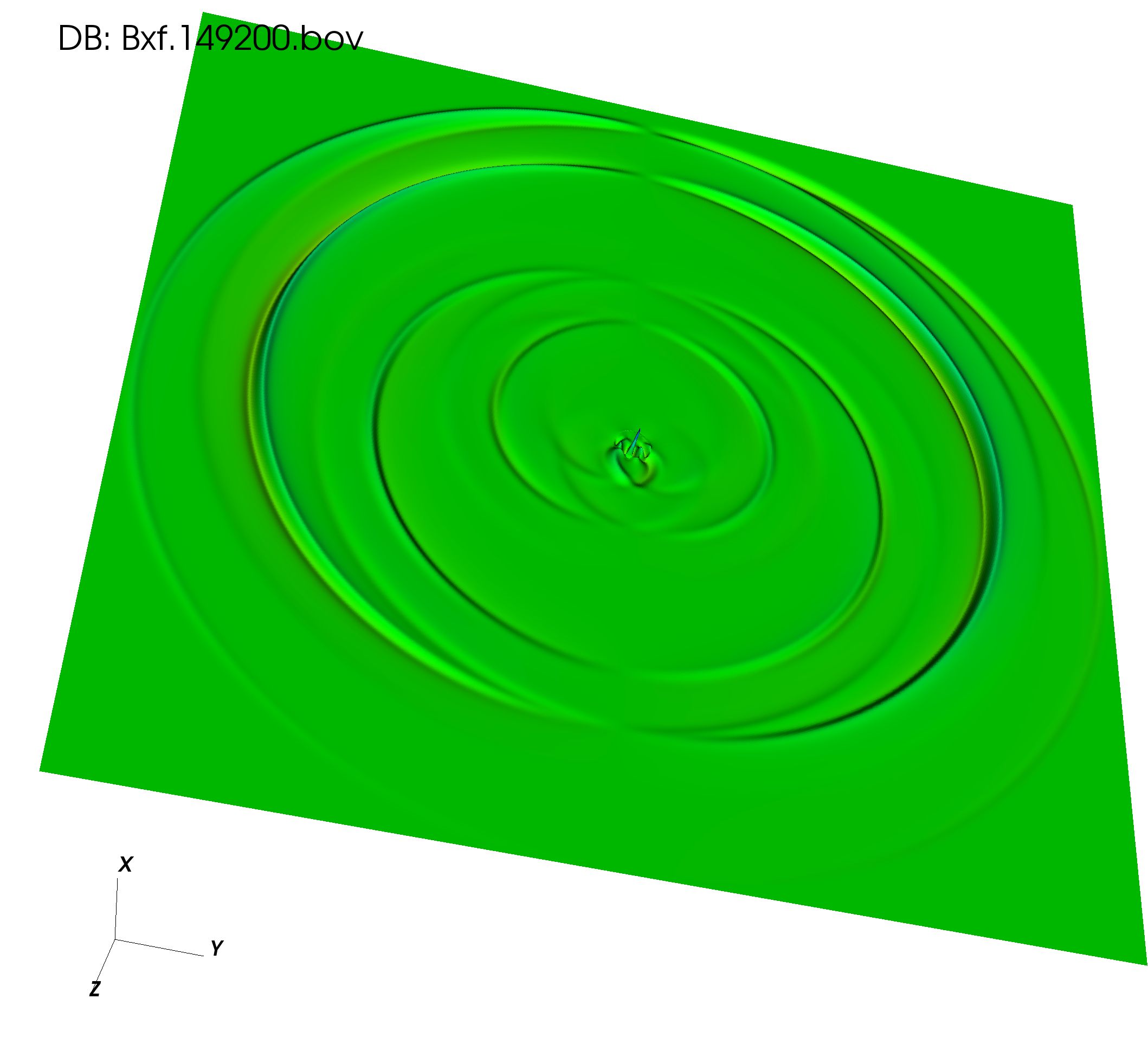}
Scattering from dielectric cylinder:  (a) t = 30k , \qquad   (b)  t = 37.8k
\caption{The self-consistently generated $B_x(x,y)$-field after the 1D incident pulse with $B_y = B_y(x)$ scatters from a local dielectric with refractive index $n(x,y)$ at times:  (a)  $t = 23.4K$, corresponding to Fig. 4a for $E_z$, and (b) $t = 49.2k$, corresponding to Fig. 5a for $E_z$.
}
\end{figure}

\subsection{Time Dependence of $\mathcal{E}(t)$ on Perturbation Parameter $\delta$}
The discrete total electromagnetic energy $\mathcal{E}(t)$. Eq. (30), is not constant since our current QLA is not totally unitary.  However
the variations in $\mathcal{E}$ decrease significantly as $\delta \rightarrow 0$.  $\delta$ is a measure of the discrete lattice spacing. The maximal variations occur shortly after the 1D pulse scatters from the 2D dielectric object.  For $\delta = 0.1$, this occurs around $t = 15k$, with variations in the 5th decimal, Eq. (31).  However, when one reduces $\delta$ by a factor of $10$ on the same lattice grid, then one recovers the same physics a factor of $10$ later in time since $\delta$ controls the speed of propagation in the vacuum.  Thus the wallclock time of a QLA run is also increased by this factor of $10$.  We find, for $\delta = 0.01$, that the largest deviation in the total electromagnetic energy is now in the 8th decimal, Eq.. (31). 
\begin{equation}
\boldsymbol{\delta = 0.1} : \begin{array}{ccc}
time & \mathcal{E}(t) \times10^{-4} \\ 
0  & 1.442426615\\
15000 &  1.4424\textcolor{red}{48...} \end{array} \quad \quad \quad
\boldsymbol{\delta = 0.01}  : \begin{array}{ccc}
time & \mathcal{E}(t) \times10^{-4}\\ 
0  & 1.442426615\\
150000 &  1.4424266\textcolor{red}{20}
\end{array}
\end{equation}
For $\delta = 10^{-3}$, there is variation in $\mathcal{E}$ in the 11th decimal.

\subsection{Multiple reflections/transmissions within dielectric cylinder}
We now examine the scattered electromagnetic fields - particularly the polarization $E_z(x,y)$ - within and in the vicinity of the
dielectric cylinder.  These plots complement the global scattered $E_z(x,y)$ in Figs. 3a, 4a and 5a, but for better resolution we choose $\delta = 0.01$ and a slightly different ratio of pulse width to dielectric cylinder diameter.  In Figs 8 - 13, the perspective is looking down from above with the 1D pulse propagating from left to right ($\rightarrow$), seen as a dark vertical band.  The dielectric cylinder appears
as a pink cylinder with the smaller darker pink being the base of the cylinder.  The time is expressed in normalized time :  $t = t^*/{10}$, where $t^*$ is the QLA time for $\delta = 0.01$.

In Fig. 8a, at time $t = 7.2k$, a part of the incident pulse has just entered the dielectric cylinder with the transmitted $E_z$ field  starting to lag behind the main 1D vacuum pulse since
$1 < n_{cyl}$.  Also the reflected part of $E_z$ emanates from the two boundary points at the sharp vacuum-dielectric boundary and has undergone a $\pi$-phase change because the incident 1D pulse is propagating from low to high refractive index.  In Fig 8b the slower transmitted $E_z$ wavefront within the dielectric is very evident, as is the reflected part of $E_z$ back into the vacuum.

 By $t = 18k$, Fig 7b, the 1D pulse has propagated past the dielectric.  The $E_z$-field within the dielectric is now being focussed due to its motion towards the backend of the cylinder, with its increasing amplitude but reduced base.   As it reaches the backend of the dielectric, part of $E_z$ will be transmitted into the vacuum while the other part will be reflected back into the dielectric - but now without any phase change since the pulse is propagating from high to low refractive index.

\begin{figure}[!b]\centering
\includegraphics[width=3.2in]{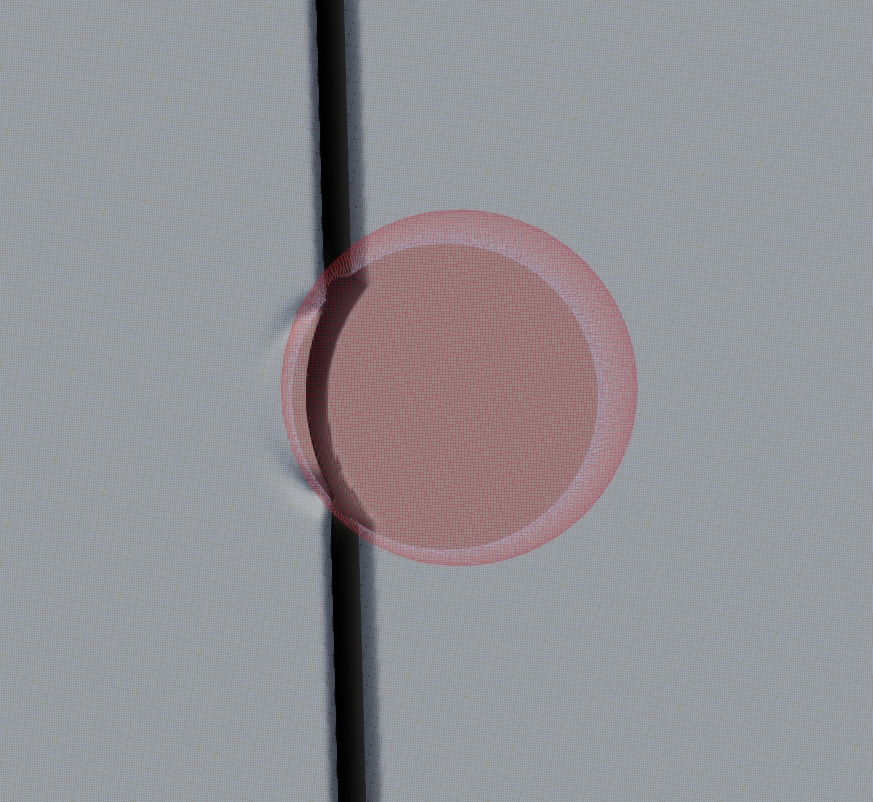}
\includegraphics[width=3.2in]{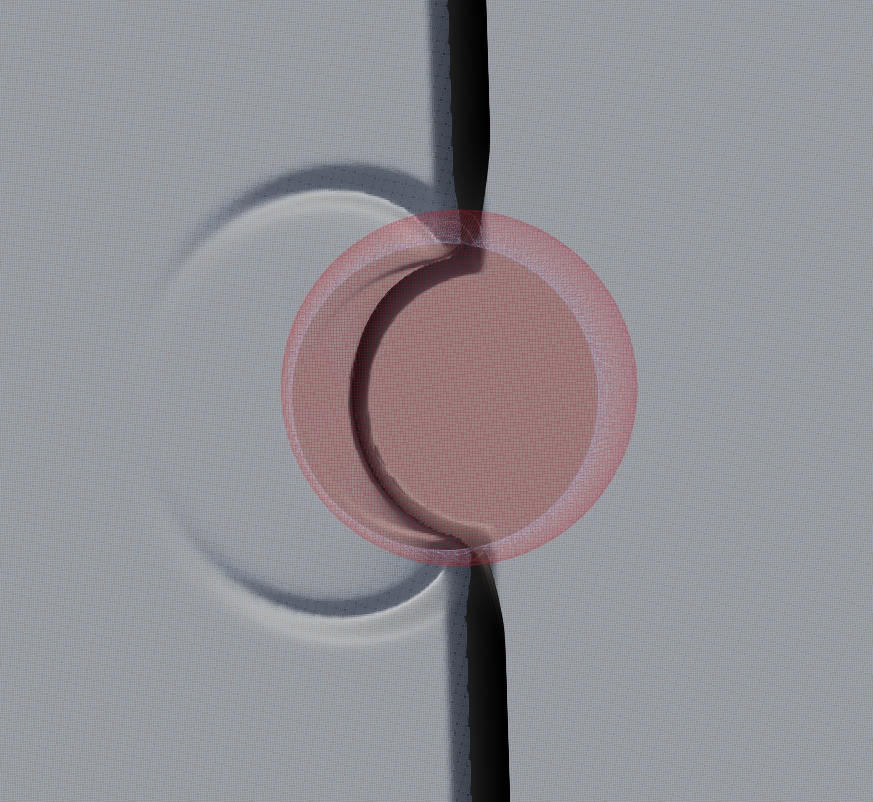}
$E_z$ field: (a)  t = 7.2k \qquad  \qquad  \qquad (b)  t = 12k
\caption{ A view from the $z$-axis of the 1D incident $E_z$ wavefront, with x-y the plane of the page.  The vacuum pulse is propagating in the $x$-direction, $\rightarrow$ (a)  The 1D incident pulse has encountered the localized dielectric cylinder, with both transmission and reflection at the thin vacuum-dielectric boundary layer.  The reflected $E_z$ circular wavefront undergoes a $\pi$-phase change.
(b) The transmitted $E_z$, within the dielectric, has lower phase speed and so lags the 1D vacuum pulse.  
}
\end{figure}
\begin{figure}[!b]\centering
\includegraphics[width=3.2in]{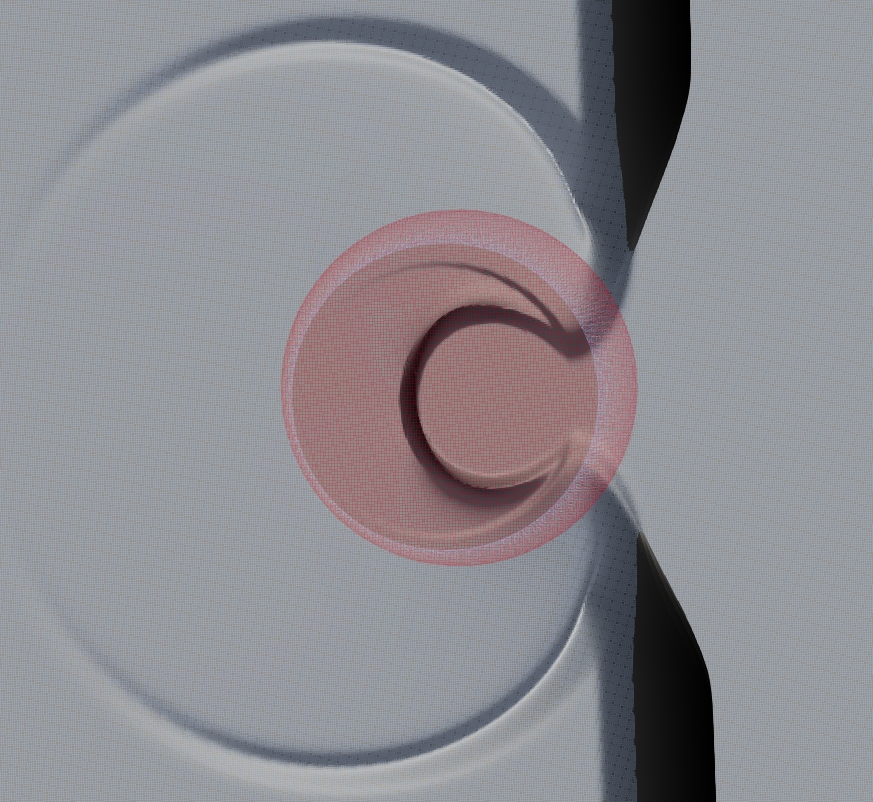}
\includegraphics[width=3.2in]{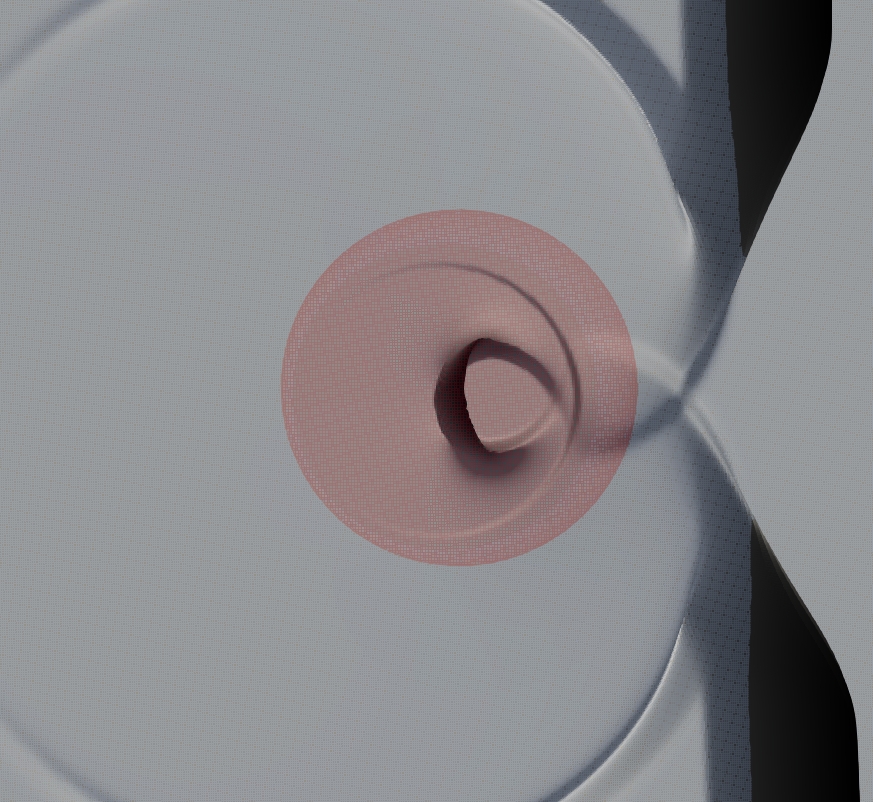}
$E_z$ wavefronts at (a) t = 18k  \qquad \qquad  \qquad  (b) t = 22.2k
\caption{ As the 1D vacuum part of the wavefront moves past the dielectric cylinder, the two vacuum-dielectric boundary "points" move closer together: (a)  at t = 18k, (b) at t = 22.2K.  The vacuum reflected wavefront  keeps radiating out.  
}
\end{figure}

\begin{figure}[!b]\centering
\includegraphics[width=3.2in]{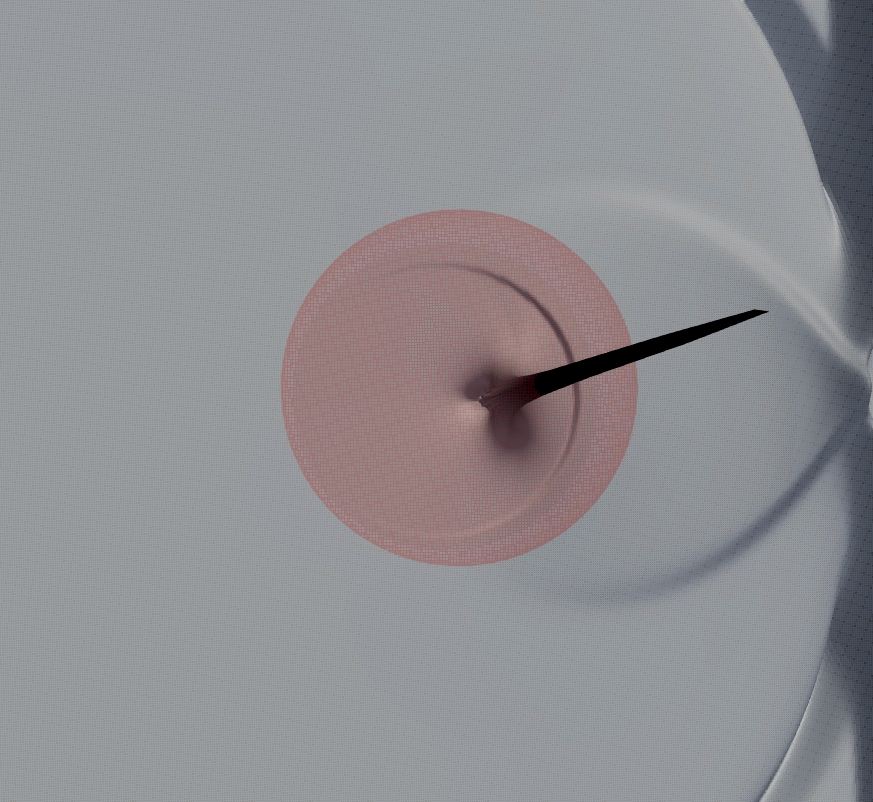}
\includegraphics[width=3.2in]{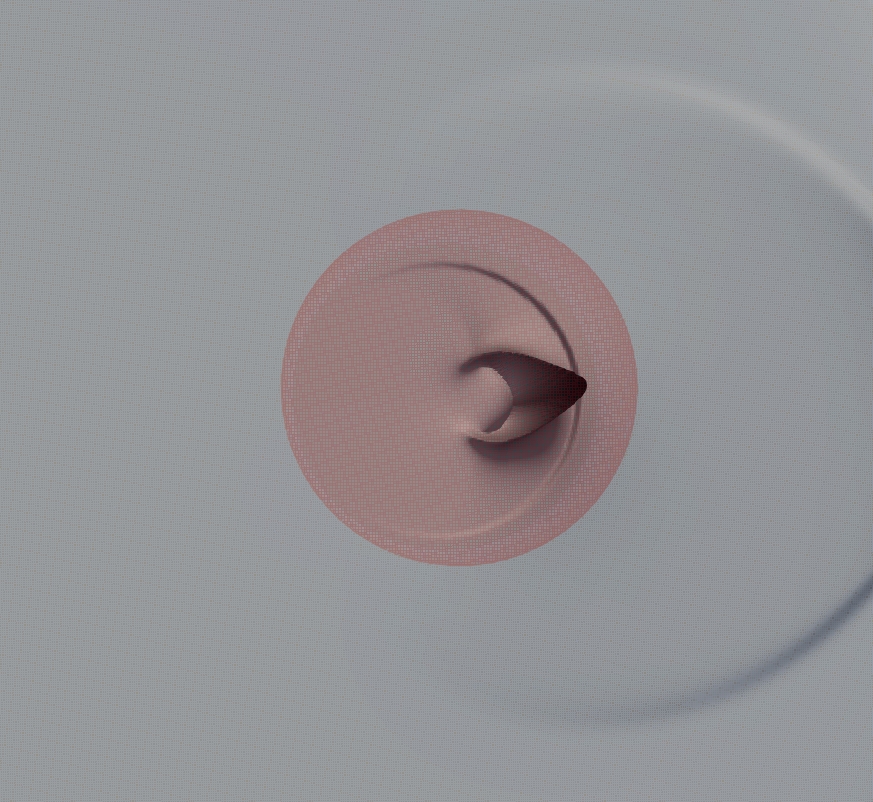}
$E_z$ wavefronts at (a) t = 28.2k  \qquad \qquad  \qquad  (b)  t = 33k
\caption{Wavefronts of $E_z$ at times (a) t = 28.2k, and (b) t = 32k around and within the dielectric cylinder after the original 1D pulse has moved past the dielectric.  (a)  The pinching of the two boundary "points" results in a focussing of $E_z$ and its subsequent spiking at $t = 28.2k$.  This spike now propagates towards the backend of the dielectric cylinder, (b), and "diffuses",  One should also note the wavefront emanating from the 1D vacuum pulse.
}
\end{figure}

\begin{figure}[!b]\centering
\includegraphics[width=3.2in]{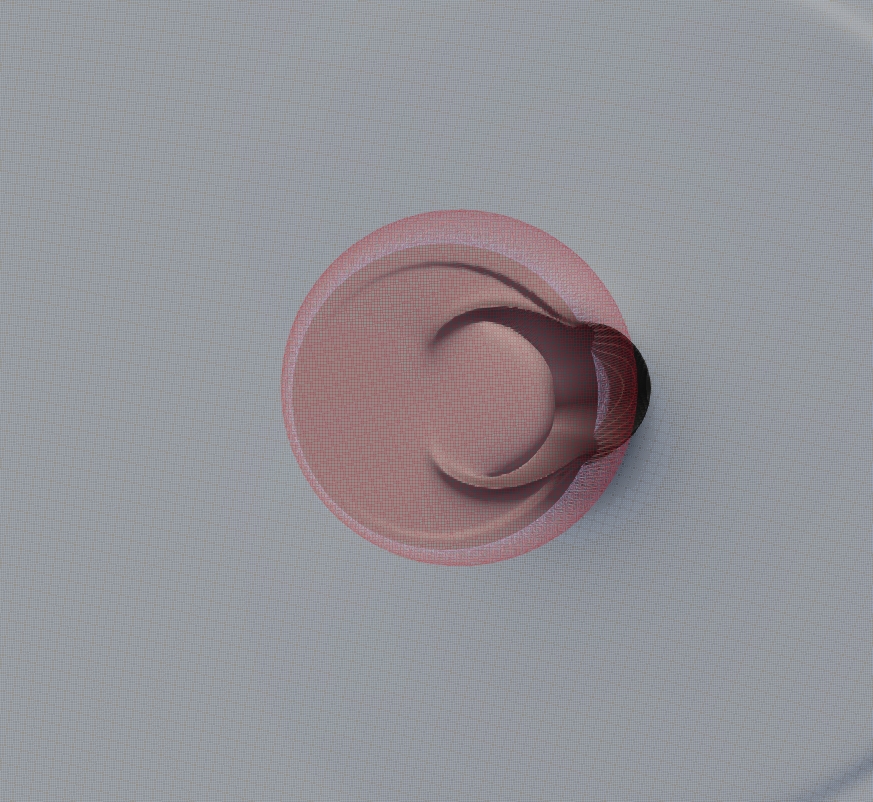}
\includegraphics[width=3.2in]{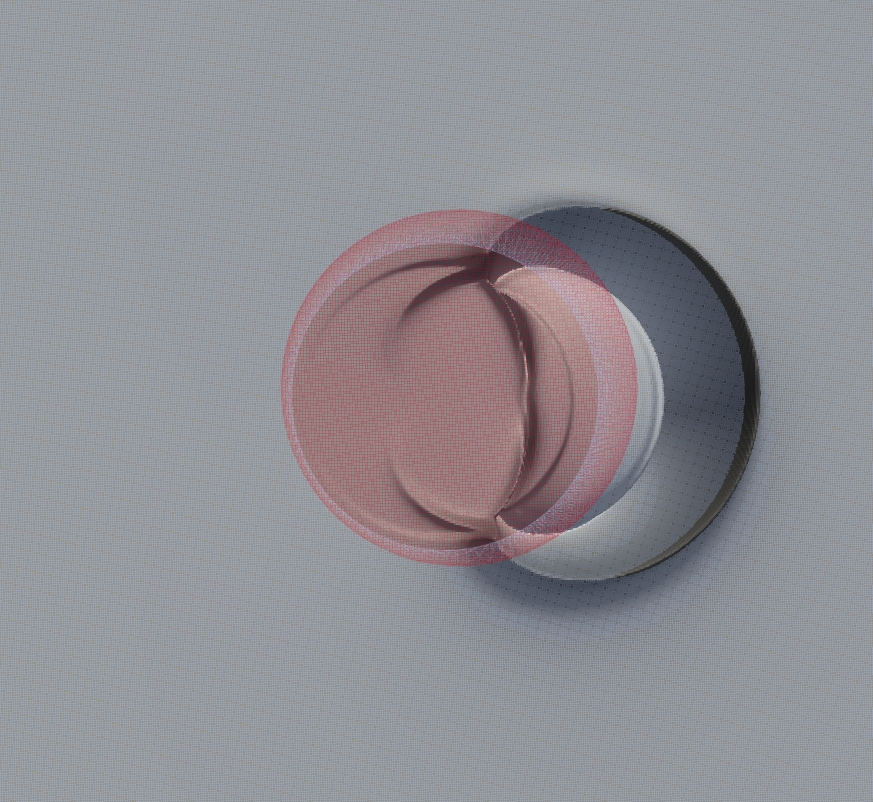}
$E_z$ wavefronts at (a) t = 38.4k  \qquad \qquad  \qquad \qquad (b)  t = 43.2k
\caption{Wavefronts of $E_z$ at times around and within the dielectric cylinder.  At (a) t = 38.4k the transmitted $E_z$ within the dielectric is radiating outward, with one part reaching the back of the dielectric and resulting in a complex transmission into the vacuum region at the back end of the dielectric, (b) at t = 43.2k, and a complex reflection back into the dielectric.  There is no phase change in the reflected $E_z$.
}
\end{figure}

\begin{figure}[!b]\centering
\includegraphics[width=3.2in]{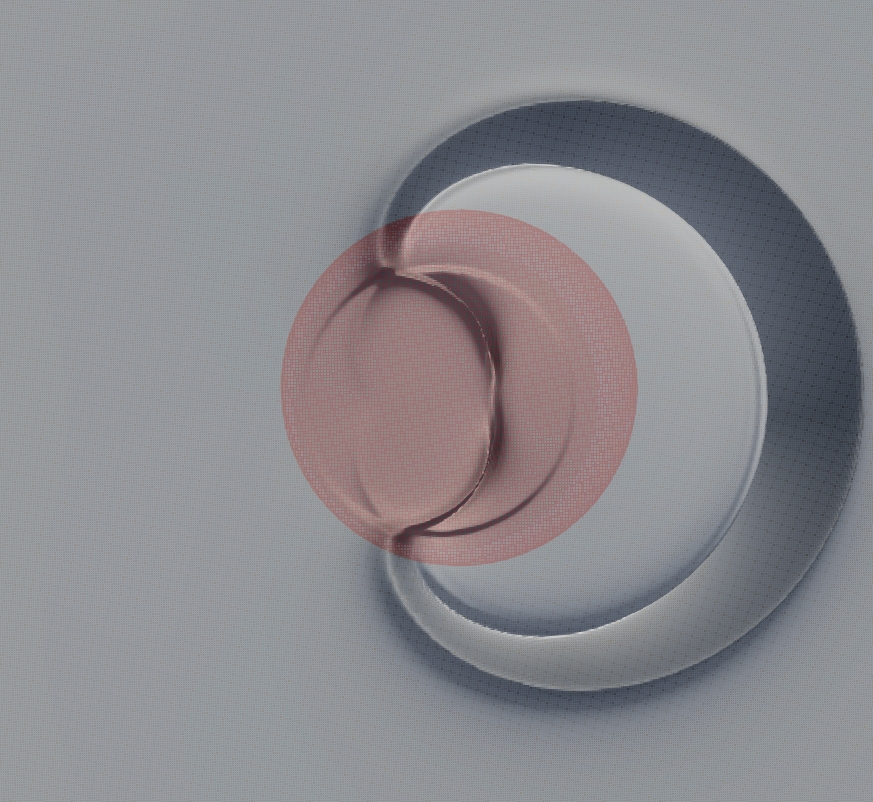}
\includegraphics[width=3.2in]{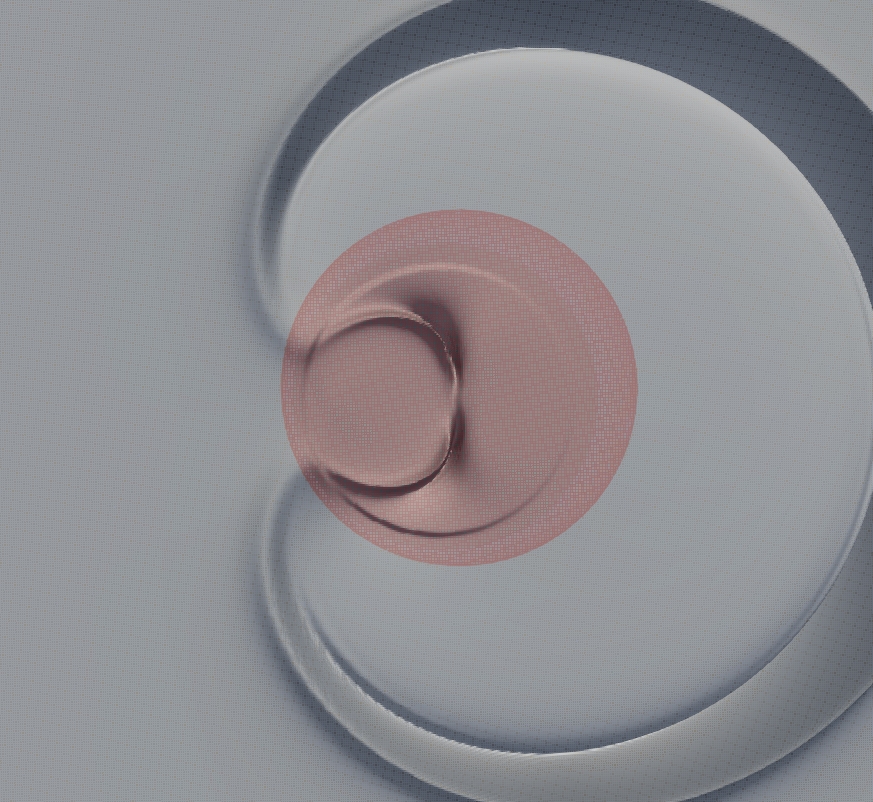}
$E_z$ wavefronts at (a) t = 47.4k  \qquad \qquad  \qquad  (b)  t = 52.2k
\caption{Wavefronts of $E_z$ at times (a) t = 47.4k, and (b) t = 52.2k around and within the dielectric cylinder.  The major vacuum wavefront that is transmitted out of the dielectric now radiates out in the xy-plane.  The two boundary contact "points" of the wavefront are now propagating back to the front of the dielectric cylinder, as clearly seen in (a) and (b).
These localized wavefronts will have their global wavefronts similar to those shown in Figs. 3a, 4a and 5a
}
\end{figure}

\begin{figure}[!b]\centering
\includegraphics[width=3.2in]{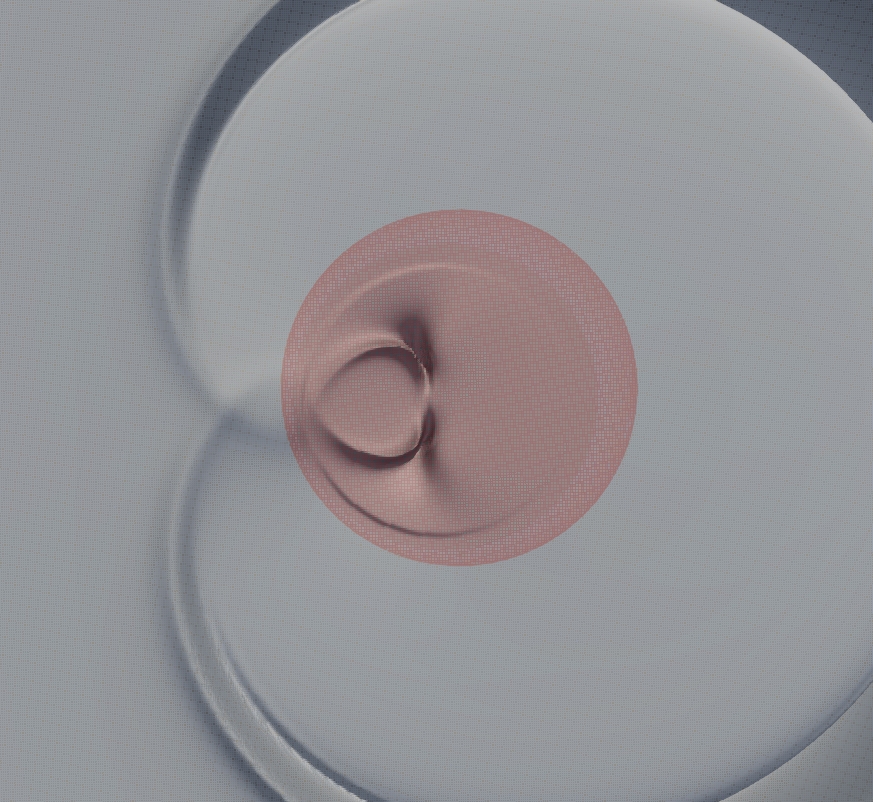}
\includegraphics[width=3.2in]{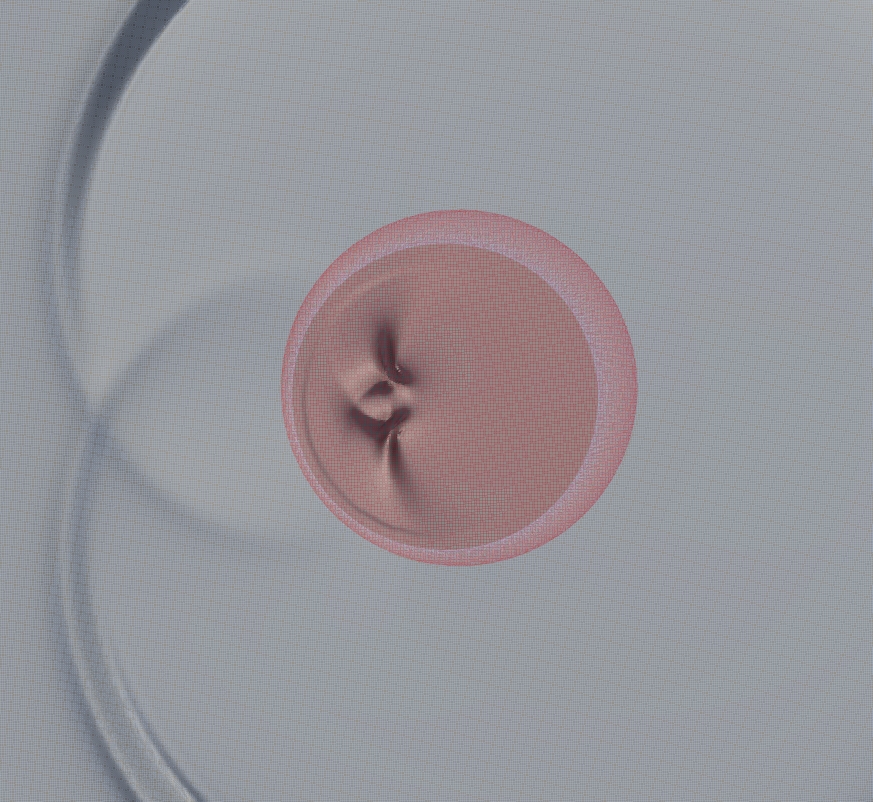}
$E_z$ wavefronts at (a) t = 55.8k  \qquad \qquad  \qquad  \qquad  (b)  t = 60k
\caption{Wavefronts of $E_z$ at times (a) t = 55.8k, and (b) t = 60k around and within the dielectric cylinder.  
}
\end{figure}

\section{Dissipative Classical Systems, Open Quantum Systems and Kraus Representation}   

So far we have treated Maxwell equations as a closed system based on the energy conservation dictated from the Hermiticity and positive definiteness of the constitutive matrix $\mathbf{W}$, Eq. (7), since we have restricted ourselves to perfect materials. However, when we wish to consider actual materials there is dissipation.  This immediately defeats any attempt to pursue a unitary representation in the original Hilbert space. The obvious question is :  can we embed our dissipative system into a higher dimension closed Hilbert space, and thus recover unitary evolution in this new space and build an appropriate QLA that can be encoded onto quantum computers? To accomplish this, 
we resort to open quantum system theory \textbf{[21]} to describe classical dissipation as the observable result of interaction between our system of interest and its environment.  

For a closed quantum system, the time evolution of a pure state $|\psi(t) \rangle$  
is given by the unitary evolution from the Schrodinger equation : 
$|\psi(t) \rangle = U(t) |\psi(0) \rangle$
with $U = exp[-i t H_0]$ unitary for the Hermitian Hamiltonian $H_0$.  The evolution of the density matrix, 
$\rho = |\psi \rangle \langle \psi |$, 
 is governed by the corresponding  von Neumann equation :  $\rho(t) = U(t) \rho(0) U^{\dagger}(t)$. The density matrix formulation is required when dealing with composite systems. Kraus realized that the density matrix retains its needed properties if one generalized its evolution operator to
\begin{equation}\label{1}
\rho(t) = \sum_k K_k \rho(0) K_k^{\dagger}   ,  \quad \text{with  } \sum_k K_k^{\dagger} K_k = I
\end{equation}
where the only restriction on the set of so-called Kraus matrices $K_k$ is that the sum of $K_k^{\dagger} K_k$ is the identity matrix. The evolution of the density matrix, Eq. (32), is no longer unitary for $k \ge 2$.

The Kraus representation \textbf{[21]} is most useful when dealing with quantum noisy operations due to interaction with an environment. For those
problems in which this noisy operation translates into a dissipative process, the Hamiltonian for the system in the Schrodinger representation has both a Hermitian part, $H_0$, and an anti-Hermitian part, $i \, H_1$, that models the dissipation. A simple but non-trivial example is the $1D$  Maxwell equations (without sources) for a homogeneous scalar medium with electrical losses,
\begin{equation}\label{2}
i \frac{\partial}{\partial t} \left[\begin{array}{cc}
E_y\\
H_z
\end{array}\right] = \left[\begin{array}{cc}
0 & \frac{\epsilon^*}{|\epsilon|^2} \hat{p}_x\\
\frac{1}{\mu_0}\hat{p}_x & 0
\end{array}\right] \left[\begin{array}{cc}
E_y\\
H_z
\end{array}\right] 
\end{equation}
with complex permittivity $\epsilon=\epsilon_R+i\epsilon_I.$  $\epsilon^*=\epsilon_R-i\epsilon_I.$ $\hat{p}_x=-i\partial_x$ is the momentum operator.
Introducing the Dyson map $\boldsymbol{\rho}=diag(|\epsilon|/\sqrt{\epsilon_R},\sqrt{\mu_0})$ into Eq. (33) and after some algebraic manipulations the evolution equation can be written as
\begin{equation}\label{3}
i \frac{\partial \mathbf{Q}}{\partial t} =\Big[v_\delta\Big(\sigma_x+\frac{1}{2}\delta\sigma_y\Big)\hat{p}_x-\frac{i}{2}\delta{v}_\delta\sigma_x\hat{p}_x\Big]\mathbf{Q},
\end{equation}
where the state vector $\mathbf{Q} = \boldsymbol{\rho} \mathbf{u}$, where $\mathbf{u}$ is defined in Eq. (6).
$\delta=\epsilon_I/\epsilon_R$ is the loss angle, $v_\delta$ is the phase velocity  $v_\delta=1/\sqrt{\epsilon_R\mu_0 (1+\delta^2)}$ 
and $\sigma_x, \sigma_y, \sigma_z$ are the Pauli matrices.

\subsection{Classical Dissipation as a Quantum Amplitude Damping Channel}   
In symbolic form, the Maxwell equations with electric resistive losses, Eq. (34), can be written in the Schrodinger-form
\begin{equation}\label{4}
i \frac{\partial | \psi_S \rangle}{\partial t} = \left( \hat{H}_0(\mathbf{r}) - i \, \hat{H}_1(\mathbf{r})\right) | \psi_S \rangle
\end{equation}
where the Hamiltonians $\hat{H}_0$ and $\hat{H}_1$ are Hermitian, and the dissipative operator $i \, \hat{H}_1$ is anti-Hermitian and positive definite.
The positive definiteness requirement for the specific case of propagation in a lossy medium translates to
\begin{equation}
Im \left[ E_y^* \frac{\partial H_z}{\partial x}  + H_z^* \frac{\partial E_y}{\partial x}    \right] > 0,   \quad \text{with}  \quad \epsilon_I > 0.
\end{equation}
In general  the dissipative operator $\hat{H}_1$ is relatively simple, and models the phenomenological or coarse-graining of the underlying
microscopic dissipative processes.

We aim to represent the dissipation in the Schrodinger picture, Eq. (35), as an open quantum system S interacting with its environment $Env$. The full system, $S+Env$, is closed, and hence its time evolution is unitary.  
Let $\hat{\mathcal{U}}$  be this unitary operator, and $\rho$ the total density matrix with $\hat{\mathcal{U}} :  \rho(0) 
\rightarrow \rho(t)$.  We make the usual assumption that the initial total density matrix is separable into the system and into the environmental Hilbert spaces : $ \rho(0) = \rho_S(0) \otimes \rho_E(0)$.  A quantum operation $\textit{E}$ on the open system of interest is defined as the map that propagates the open system density in time $t$:
\begin{equation}
\rho_S(t) = \textit{E} (\rho_S(0))  .
\end{equation}
But under the conditions of  initial separability, the action of the full unitary operators on the total density matrix will yield, after taking the trace over the environment,
\begin{equation}
\rho_S(t) =Tr_{\textit{E}} \left( \rho(t) \right)  = 
Tr_{\textit{E}} \left( \hat{\mathcal{U}} \, \rho(0) \hat{\mathcal{U}}^{\dagger} \right) 
\end{equation}
Assuming a stationary environment, $\rho_E(0) = | a \rangle \langle a |$, Eq. (38) can be written
\begin{equation}
\rho_S(t) = \sum_{\mu} \hat{K}_\mu \, \rho_S(0) \hat{K}_{\mu}^{\dagger} .
\end{equation}
where $\hat{K}_\mu = \langle \mu | \hat{\mathcal{U}}  | a \rangle$.
These operators $\hat{K}_\mu$ will form a Kraus representation for the quantum operation 
\textit{E} for an open system, Eq. (37), provided the so-called Kraus operators satisfy
the extended "unitarity" condition
\begin{equation}
\sum_\mu K_\mu^{\dagger} K_\mu = I
\end{equation}
Note that the individual Kraus operator need not be unitary.
Based on this framework for open quantum systems we proceed to construct a physical unitary dilation for the combined system-environment by identifying dissipation as an amplitude damping operation, \textbf{[21]}.

Let $d$ be the dimension of the system Hilbert space, and $r$ the dimension of the dissipative Hamiltonian $H_1$, Eq. (35).  We require $d \ge 2r$, for optimal results but the dilation technique can be also applied to systems with $d=r$. If the system was quantum mechanical in nature, then there can be a set of $d^2$ Kraus operators at most.
The matrix representation of the total unitary dilation evolution operator consists of listing all the Kraus matrices in the first column block.
The remaining columns must then be determined so that $\hat{\mathcal{U}}$ is unitary.  This unitary dilation is
equivalent to the Stinespring dilation theorem \textbf{[25]}.  The advantage of the Kraus approach is that it avoids the need to 
actually know the physical properties of the environment.

Returning to the Schrodinger representation of the classical system Eq. (35), one can employ the Trotter-Suzuki expansion to $exp[- i \delta t (\hat{H}_0 - i \hat{H}_1)]$
\begin{equation}\label{9}
| \psi(\delta t) \rangle = \left[ e^{-i \delta t \hat{H}_0} \cdot e^{- \delta t \hat{H}_1} +\textit{O}(\delta t^2) \right] | \psi(0) \rangle .
\end{equation}
Even though $exp(-\delta t \hat{H}_1)$ is not unitary, $\hat{H}_1$ is Hermitian and can be diagonalized by a unitary transformation $U_1$
\begin{equation}
  \hat{H}_1 = \hat{U}_1 \hat{D}_1 \hat{U}_1^{\dagger}   \quad \text{with diagonal} \quad \hat{D}_1 = 
  diag[\gamma_1 , ... ,\gamma_r ],
\end{equation}
where $\gamma_i > 0 $  are the dissipative rate eigenvalues of $\hat{H}_1$.  Thus Eq. (41) becomes
\begin{equation}
| \psi(\delta t) \rangle =  \left[ e^{-i \delta t \hat{H}_0}  \hat{U}_1 \hat{K}_0 \hat{U}_1^{\dagger}  +\textit{O}(\delta t^2) \right] | \psi(0) \rangle,
\end{equation}
where $\hat{K}_0$ is
\begin{equation}
\hat{K}_0 =\left[\begin{array}{cc}
\hat{\boldsymbol{\Gamma}}_{r \times r} &  0_{r \times r} \\
0_{(d-r) \times (d-r)} & I_{(d-r) \times (d-r)}
\end{array}\right]  , \quad \text{with diagonal  } 
\hat{\boldsymbol{\Gamma}}_{r \times r} = diag[e^{-\gamma_1 \delta t} ... e^{-\gamma_r \delta t}].
\end{equation}
The non-unitary $\hat{K}_0$ will be one of our Kraus operators, and it describes the physical dissipation in the open system.
We must now introduce a second Kraus operator $\hat{K}_1$ so that  $\hat{K}_0^{\dagger} \hat{K}_0 +
\hat{K}_1^{\dagger} \hat{K}_1  = \mathcal{I}$ :
\begin{equation}
\hat{K}_1 =\left[\begin{array}{cc}
0_{(d-r) \times (d-r)}&  0_{(d-r) \times (d-r)} \\
\sqrt{I_{r \times r} - \hat{\boldsymbol{\Gamma}}^2} & 0_{r \times r}
\end{array}\right]  .
\end{equation}
$\hat{K}_1$ represents a transition that is not of direct interest.  These Kraus operators $\hat{K}_0 , \hat{K}_1$ are the multidimensional analogs of the quantum amplitude damping 
channel \textbf{[21]}:  with $\hat{K}_0$ corresponding to the dissipation processes, while $\hat{K}_1$ corresponds to an unwanted quantum transition.  

The block structure of the final \textit{unitary} dilation evolution operator $\hat{\mathcal{U}}_{diss}$, corresponding to the non-unitary dissipation operator 
$e^{-\delta t \hat{H}_1}$, consists of column blocks of the Kraus operators $( K_0 \quad K_1 ...)^T$, and the remaining column blocks are of those
matrices required to make $\hat{\mathcal{U}}_{diss}$ unitary \textbf{[21]}:
\begin{equation}
\hat{\mathcal{U}}_{diss}= \left[\begin{array}{cccc}
 \hat{\boldsymbol{\Gamma}}& 0 & 0 & -\sqrt{I_{r \times r} - \hat{\boldsymbol{\Gamma}}^2}  \\
0 & I_{(d-r) \times (d-r)}  & 0 & 0 \\
0 & 0 & I_{(d-r) \times (d-r)} &0 \\
\sqrt{I_{r \times r} - \hat{\boldsymbol{\Gamma}}^2}  & 0 & 0& \hat{\boldsymbol{\Gamma}}
\end{array}\right] .
\end{equation}
Thus, it can be shown that the evolution of the system $| \psi_0 \rangle$ and environment $| 0 \rangle$ 
is given by
\begin{equation}\label{13}
 | 0 \rangle | \psi_0 \rangle = 
\frac{1}{\sqrt{E_0}} \sum_q \psi_{0q} | 0 \rangle | q \rangle  \rightarrow
  | 0 \rangle \otimes e^{-i \delta t \hat{H}_0}  \hat{U}_1 \hat{K}_0 \hat{U}_1^{\dagger} | \psi_0 \rangle  +
  |1 \rangle \otimes \hat{K}_1 | \psi_0 \rangle ,
\end{equation}
where measurement of the first qubit-environment by $| 0 \rangle \langle 0 | \otimes I_{d \times d}$ yields a state analogous to 
$ | 0 \rangle | \psi(\delta t) \rangle $.  Finally, on taking the trace over the environment will yield the desired system state $|\psi(\delta t) \rangle$
The corresponding quantum circuit for Eq. (47) is shown in Fig. 14.\\
\begin{figure}[!b]\centering
\includegraphics[width=4.3in]{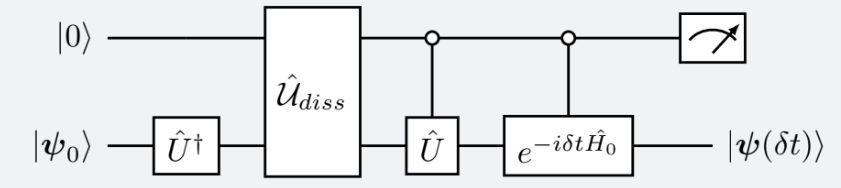}
\caption{Quantum circuit diagram for Eq. (47).  
}
\end{figure}

It is important to highlight that the implementation of the dissipative case is directly overlapping with the QLA framework. The QLA can be used to implement the $\exp{[-i\delta t\hat{H}_0]}$ part in Eq. (43) as proposed in the previous sections. Specifically for the lossy medium, the exponential operator of the  $\hat{H}_0$ term in Eq. (35) can be easily handled with QLA \textbf{[12,13]}.

\section{Conclusions}
The Schrodinger-Dirac equations are the backbone of the work presented here on Maxwell equations both
in lossless inhomogeneous  and lossy dielectric media. In both cases, straightforward application of unitary algorithms fail :  in the first case, somewhat surprisingly
one finds that even though a Dyson map points to the required electromagnetic field variables in a tensor dielectric, its implementation has till now defied a fully unitary representation.   Our current QLA approach  requires some external non-unitary operators that recover the terms involving the spatial derivatives on the refractive indices of the medium.  These sparse matrices can be modeled by the sum of a linear combination of unitaries (LCU), which can then be encoded onto a quantum computer \textbf{[23, 24]}.  In the second case, handling dissipative systems immediately forces us to consider an open quantum system interacting with its environment.  Typically this forces us into a density matrix formulation and a clever introduction of what is known as Kraus operators \textbf{[21, 25]}.  The beauty of the Kraus representation is that even though the system of interest is interacting with the environment, the Kraus operators do not need detailed information on the environment.  

We presented detailed 2D scattering of a 1D electromagnetic pulse off localized dielectric objects.  QLA is an initial value scheme.  No internal boundary conditions are imposed at the vacuum-dielectrix interface.
 For dielectrics will large spatial gradients in the refractive index, QLA simulations show
 strong internal reflection/transmission within the dielectric object.  These lead to quite complex time evolution of wavefronts from the dielectric objects.  On the other hand, for weak spatial gradients in the refractive index there are negligible reflections from the vacuum-dielectric interface.  This is reminiscent of WKB-like effects in the ray tracing approximation.

In considering  the dissipative counterpart, one must now include both the system and its environment in order to get a closed system with unitary representation.  The Kraus operators are the most general scheme 
that will retain the properties of the density matrix in time.  The probability of obtaining the desired non-unitary evolution of the open system after the measurement operator $\hat{P}_0 = | 0 \rangle \langle 0 | \otimes 
I_{d \times d} $ is
\begin{equation}
p(0) = \sum_{i=1}^r e^{-2 \gamma_i \delta t} | \psi_{i0}|^2 +\sum_{i=r+1}^d |\psi_{i0}|^2 \ge 1+(e^{-2\gamma_{max}\delta t}-1)\sum_{i=1}^r e^{-2 \gamma_i \delta t} | \psi_{i0}|^2
\end{equation}
The form of the unitary operator $\hat{\mathcal{U}}_{diss}$, Eq. (42) implies that it can be decomposed into r two-level unitary rotations $\hat{R}_y(\theta_i)$ with $\cos(\theta_i/2) = e^{- \gamma_i \delta t}.$ Then, the quantum circuit implementation of $\hat{\mathcal{U}}_{diss}$ requires $O(r log_2^2 d)$ CNOT - and $\hat{R}_y(\theta_i)$ quantum gates, so that there is a improvement in the circuit depth  of $O(r/d^2)$.  Our multidimensional amplitude damping channel approach is directly related to the 
Sz. Nagy dilation by a rotation.  The Sz. Nagy dilation \textbf{[26]} is the minimal unitary dilation containing the original dissipative (non-unitary) system.

 \section*{Acknowledgments}
This research was partially supported by Department of Energy grants DE-SC0021647, DE-FG02-91ER-54109, DE-SC0021651, DE-SC0021857,   and DE-SC0021653.  This work has been carried out within the framework of
the EUROfusion Consortium, funded by the European Union via the Euratom Research and Training Programme (Grant Agreement No. 101052200 - EUROfusion). Views and opinions expressed, however, are those of the authors only and do not necessarily reflect those of the European Union or the European Commission. Neither the European Union nor the European Commission can be held responsible for them. E. K. is supported  by the Basic Research Program, NTUA, PEVE.  K.H is
supported by the National Program for Controlled Thermonuclear Fusion, Hellenic Republic.
This research used resources of the National Energy Research Scientific Computing Center (NERSC), a U.S. Department of Energy Office of Science User Facility located at Lawrence Berkeley National Laboratory, operated under Contract No. DE-AC02-05CH11231 using NERSC award FES-ERCAP0020430.

\section{References}
[1]  B. M. Boghosian \& W. Taylor IV, "Quantum lattice gas models for the many-body Schrodinger equation", Internat. J. Modern Phys. \textbf{C8}, 705-716 (1997).

[2]  [1]  B. M. Boghosian \& W. Taylor IV, "Simulating quantum mechanics on a quantum computer", Physica \textbf{D 120}, 30-42 (1998).

[3]  J. Yepez \& B.M. Boghosian, "An efficient and accurate quantum lattice-gas model for the many-body Schroedinger wave equation," Computer Physics Communications, 146, (2002) 280-294.

[4]  G. Vahala, J. Yepez \& L. Vahala, "Quantum lattice gas representation of some classical solitons", Phys. Lett  \textbf{A 310}, 187-196 (2003)

[5] G. Vahala, L. Vahala \& J. Yepez, "Inelastic vector soliton collisions:  a lattice-based quantum representation", Pbhil. Trans:  Math, Phys. and Eng. Sciences, Roy. Soc. \textbf{362}, 1677-1690 (2004)

[6]  J. Yepez, G. Vahala, L. Vahala \& M. Soe, "Superfluid turbulence from quantum Kelvin waves to classical Kolmogorov cascades", Phys. Rev. Lett. \textbf{103}, 084501 (2009)

[7]  G. Vahala, J. Yepez, L. Vahala, M. Soe, B. Zhang \& S. Ziegeler, "Poincare recurrence and spectral cascades in three-dimensional quantum turbulence", Phys. Rev. \textbf{E 84}, 046713 (2011)

[8]  B. Zhang, G. Vahala, L. Vahala, \&M. Soe, "Unitary quantum lattice algorithm for two dimensional quantum turbulence", Phys. Rev. \textbf{E 84}, 046701 (2011)

[9]  G. Vahala, B. Zhang, J. Yepez, L. Vahala \& M. Soe, "Unitary qubit lattice gas representation of 2D and 3D quantum turbulence", in \textit{Advanced Fluid Dynamics}, ed. H. W. Oh, (InTech , 2012), Chpt 11., p. 239 - 272.

[10]  L. Vahala, M. Soe, G. Vahala \& J. Yepez, "Unitary qubit lattice algorithms for spin-1 Bose Einstein condensates",  Rad. Effects Def. Solids \textbf{174}, 46-55 (2019).

[11]  G. Vahala, L. Vahala \& M.Soe, "Qubit unitary lattice algorithms for spin-2 Bose-Einstein Condensates,I - Theory and Pade Initial Conditions", Rad. Effects Def. Solids \textbf{175}, 102-112 (2020).

[12] G. Vahala, M. Soe \& L. Vahala, "Qubit unitary lattice algorithms for spin-2 Bose-Einstein Condensates,II - vortex reconnection simulations and non-Abelian vortices", Rad. Effects Def. Solids \textbf{175}, 113-119 (2020).

[13]  S. Palpacelli \& S. Succi "The Quantum Lattice Boltzmann Equation:  Recent Developments", Comm. Computational Phys. \textbf{4}, 980-1007 (2008).

[14]  S. Succi, F. Fillion-Gourdeau, \& S. Palpacelli, "Quantum lattice Boltzmann is a quantum walk", EPJ Quantum. Technol \textbf{2}, 12 (2015).

[15]  G. Vahala, L. Vahala, M. Soe, \& A. K. Ram, "Unitary quantum lattice simulations for Maxwell equations in vacuum and in dielectric media", J. Plasma Phys. \textbf{86}, 905860518 (2020).

[16]  G. Vahala, J. Hawthorne, L. Vahala, A. K. Ram \& M. Soe, "Quantum lattice representation for the curl equations of Maxwell equations", Rad. Effects and Defects in Solids \textbf{177}, 85 (2022).

[17]  G. Vahala, M. Soe, L.Vahala, A. Ram, E. Koukoutsis, \& K. Hizanidis, "Qubit Lattice Algorithm Simulations of Maxwell's Equations for Scattering from Anisotropic Dielectric Objects", e-print arXiv:2301.13601 (2023).

[18]  A. Oganesov, G. Vahala, L. Vahala \& M. Soe, "Effect of Fourier transform on the streaming in quantum lattice gas algorithms", Rad. Effects and Defects in Solids. \textbf{173}, 169 (2018).

[19]  S. A. Khan \& R. Jagannathan, "A new matrix representation of the Maxwell equations based on the Riemann-Silberstein-Weber vector for a linear inhomogeneous medium", arXiv:2205.09907v2.

[20]  E. Koukoutsis, K. Hizanidis, A. K. Ram \& G. Vahala, "Dyson maps and unitary evolution for Maxwell equations in tensor dielectric media",  Phys. Rev. \textbf{A 107}, 042215 (2023).

[21]  M. A. Nielsen \& I. L. Chuang, \textit{Quantum Computation and Quantum Information}, Cambridge University Press, 2010 (10th Ed.)


[22]  M. Suzuki, "Generalized Trotter's formula and systematic approximants of exponential operators and inner derivatives with applications to many-body probloems, Commun. Math. Phys. \textbf{51},183 (1976)

[23]  A. M. Childs \& N. Wiebe, "Hamiltonian Simulation Using Linear Combinations of Unitary Operations", Quantum Inf. and Comp. \textbf{12} , 901 (2012).

[24]  A. M. Childs, R. Kothari \& R. D. Somma, "Quantum Algorithm for Systems of Linear Equations with Exponentially Improved Dependence on Precision", SIAM J. on Comp. \textbf{46}, 1920 (2017).

[25]  W. F. Stinespring, "Positive Functions on $C^*$-Algebras", Proc. Amer. Math. Soc. \textbf{6}, 211 (1955).


[26]  V. Paulsen, \textit{Completely Bounded Maps and Operator Algebras}, Cambridge Univ. Press, 2003.

\end{document}